\documentclass[useAMS,usenatbib,usegraphicx]{mn2e}
\usepackage{amssymb}
\usepackage{amsmath}
\usepackage{booktabs}
\usepackage{hyperref}
\usepackage{multirow}
\usepackage{times}

\title[Timing analysis of PG\,1211+143]
  {X-ray timing analysis of the quasar PG\,1211+143}
\author[A.P. Lobban et al.]
  {A.P.~Lobban$^1$\thanks{e-mail: \href{mailto:al290@le.ac.uk}{al290@le.ac.uk}},
  S.~Vaughan$^1$, K.~Pounds$^1$, J.N.~Reeves$^2$ \\
  $^1$University of Leicester, X-Ray and Observational Astronomy Group, Department of Physics and Astronomy, Leicester, LE1 7RH, U.K. \\
  $^2$Astrophysics Group, School of Physical and Geographical Sciences, Keele University, Keele, Staffordshire, ST5 5BG, U.K.}
\date{\today; Accepted for publication in MNRAS}

\pagerange{\pageref{firstpage}--\pageref{lastpage}} \pubyear{2015}

\def\LaTeX{L\kern-.36em\raise.3ex\hbox{a}\kern-.15em
    T\kern-.1667em\lower.7ex\hbox{E}\kern-.125emX}

\begin{document}

\label{firstpage}

\maketitle

\begin{abstract}

We report on a timing analysis of a new $\sim$630\,ks {\it XMM-Newton} observation of the quasar, PG\,1211+143.  We find a well-defined X-ray power spectrum with a well-detected bend at $\sim$7 $\times 10^{-5}$\,Hz, consistent with the established $t_{\rm bend}$--$M_{\rm BH}$ correlation for luminous, accreting black holes.  We find the linear rms-flux relation commonly observed in accreting black hole systems and investigate the energy-dependence of the rms.  The fractional rms is roughly constant with energy on short timescales ($< 1$\,day; within observations) whereas there is enhanced soft band variability on long timescales (between observations typically spaced by a few days).  Additionally, we also report on the optical--UV variability using the OM on-board {\it XMM-Newton} and a $\sim$2-month-long overlapping monitoring programme with {\it Swift}.  We find that, although there is little UV variability within observations ($<1$\,day), UV variations of a few per cent exist on time-scales of $\sim$days--weeks.

\end{abstract}


\section{Introduction} \label{sec:introduction}

Both luminous active galactic nuclei (AGN) and black hole X-ray binaries (XRBs) are thought to be powered by accretion on to black holes (AGN: $M_{\rm BH} \sim 10^{6-9}$\,$M_{\odot}$; XRBs: $M_{\rm BH} \sim 10$\,$M_{\odot}$).  Both types of system are powerful sources of X-rays which most likely originate close to the black hole itself.  The X-ray spectra of unobscured AGN are routinely observed to consist of an array of spectral features.  In particular, they are usually dominated by a strong power-law component, thought to be produced when ultraviolet (UV) photons, emitted from an optically thick, geometrically thin accretion disc \citep{ShakuraSunyaev73}, are inverse-Compton scattered by a `corona' of hot electrons \citep{HaardtMaraschi93}.  Additional spectral features typically include a soft excess $< 2$\,keV \citep{ScottStewartMateos12}, a `Compton reflection' component $> 10$\,keV and emission lines, the strongest of which is often Fe\,K$\alpha$ fluorescence at $\sim$6.4\,keV \citep{GeorgeFabian91}.

If the dynamics of the accretion flows around black holes are governed by strong gravity, many fundamental aspects of accretion onto both supermassive black holes and stellar-mass black holes should be the same (i.e. `scale invariance').  The similarities observed in the X-ray variability of AGN and XRBs through recent observations with, for example, {\it RXTE} and {\it XMM-Newton} has lent support to the idea of `black hole unification' (e.g. \citealt{McHardy06,Fender07}).

Of particular interest are frequency-dependent measurements of the variability such as the power spectral density (PSD) and X-ray time delays between different energy bands (i.e. `X-ray time lags')\footnote{We explore the X-ray time lags in PG\,1211+143 separately in \citet{Lobban15}.} which are routinely observed to display similar behaviour but with their frequencies and amplitudes roughly scaled inversely with black hole mass (e.g. \citealt{Lawrence87}; \citealt{UttleyMcHardyPapadakis02}; \citealt{Vaughan03}; \citealt{Markowitz03}; \citealt{McHardy04}; \citealt{McHardy06}; \citealt{Arevalo08}).  Additionally, the rms amplitude of X-ray variability is commonly observed to scale linearly with the flux of the source (i.e. the `rms-flux relation').  This property of X-ray variability is observed to hold over a wide range of timescales and is common to both AGN and XRBs (e.g. \citealt{UttleyMcHardy01}; \citealt{Vaughan03}; \citealt{Gaskell04}; \citealt{UttleyMcHardyVaughan05}).

The energy output of Seyfert galaxies typically peaks at ultraviolet (UV) wavelengths.  This emission is generally considered to originate from thermal processes in the inner regions of the accretion disc.  The UV emission from many AGN is observed to be strongly variable (e.g. \citealt{Collin01}), although not as rapidly variable as the X-rays \citep{MushotzkyDonePounds93}.  If the regions responsible for X-ray/UV emission are modulated by the local accretion rate, one may expect their variations to be somewhat correlated with the two favoured coupling mechanisms consisting of Compton up-scattering of UV photons to X-rays in a hot corona (e.g. \citealt{HaardtMaraschi91}) and thermal reprocessing in the disc of primary X-rays (\citealt{GuilbertRees88,Collier98,CackettHorneWinkler07}.  In both cases, the delays in the variability are governed by the light-crossing time between the two sites. 

In order to understand the connection between different emission mechanisms, studies of inter-band correlations are important and UV/X-ray correlations have been seen in a number of AGN (e.g. \citealt{Nandra98}, \citealt{Cameron12}).  In addition, X-ray/optical correlations have been observed on longer timescales and are typically discussed in terms of a disc reprocessing model (e.g. \citealt{Arevalo08}; \citealt{Breedt09}; \citealt{Arevalo09}; \citealt{Breedt10}; \citealt{AlstonVaughanUttley13}; \citealt{McHardy14}; \citealt{Edelson15}).

Here, we apply timing techniques to PG\,1211+143, a luminous narrow-line Seyfert galaxy / quasar at a redshift of $z = 0.0809$ \citep{Marziani96} which is optically bright with a strong ``Big Blue Bump" and also X-ray bright with a typical X-ray luminosity of $\sim$10$^{44}$\,erg\,s$^{-1}$ ($L_{\rm bol} \sim 5 \times 10^{45}$\,erg\,s$^{-1}$; \citealt{Pounds03}).  The source is well-known for its spectral complexity.  In particular, it is the archetypal example example of an AGN exhibiting an ultra-fast outflow whose mass flux and energetics may be comparable to the mass accretion rate and bolometric luminosity and so important for galaxy feedback \citep{Pounds03,PoundsReeves09}.  Observations with {\it XMM-Newton} have shown the outflow to have a sub-relativistic velocity lying in the region of $v \sim $0.09--0.14$c$ \citep{Pounds03,PoundsPage06,Pounds15a}.  Here, we investigate the power spectrum, the rms spectrum and optical/UV/X-ray correlations through a new $\sim$630\,ks {\it XMM-Newton} plus $\sim$2-month-long {\it Swift} campaign of PG\,1211+143.

\section{Observations and data reduction} \label{sec:observations_and_data_reduction}

\subsection{{\it XMM-Newton}}

PG\,1211+143 was observed seven times in 2014 with {\it XMM-Newton} (\citealt{Jansen01}) between 2014-06-02 and 2014-07-07.  Each observation had a typical duration of $\sim$100\,ks, except for the fifth observation (rev2664) which was shorter ($\sim$55\,ks), with a total duration of $\sim$630\,ks.  Here, we utilize data from all instruments on-board {\it XMM-Newton}, although we only make very brief use of the Reflection Grating Spectrometer (RGS; \citealt{denHerder01}) - instead, those data are described and presented in \citet{Pounds15b} and additional companion papers.  We processed all raw data using version 14.0 of the {\it XMM-Newton} Scientific Analysis Software (\textsc{sas}\footnote{\url{http://xmm.esac.esa.int/sas/}}) package, following standard procedures.

\begin{table}
\centering
\begin{tabular}{l c c c c}
\toprule
Date / & \multirow{2}{*}{EPIC} & Total Duration & \multirow{2}{*}{Count} & \multirow{3}{*}{Flux} \\
ObsID / & \multirow{2}{*}{Camera} & (Filtered Duration) & \multirow{2}{*}{Rate} & \\
(Revolution) & & [Net Exposure] & & \\
\midrule
2014-06-02 & pn & 83 (83) [77] & 3.97 & 1.10 \\
0745110101 & MOS\,1 & 78 (78) [76] & 0.94 & 1.16 \\
(rev2652)  & MOS\,2 & 85 (85) [83] & 0.93 & 1.17 \\
\midrule
2014-06-15 & pn & 100 (98) [86] & 2.65 & 0.83 \\
0745110201 & MOS\,1 & 102 (100) [97] & 0.57 & 0.83 \\
(rev2659)  & MOS\,2 & 102 (99) [98] & 0.63 & 0.85 \\
\midrule
2014-06-19 & pn & 99 (51) [48] & 3.31 & 0.94 \\
0745110301 & MOS\,1 & 101 (51) [50] & 0.63 & 0.99 \\
(rev2661)  & MOS\,2 & 101 (51) [50] & 0.76 & 1.01 \\
\midrule
2014-06-23 & pn & 96 (91) [87] & 3.89 & 1.09 \\
0745110401 & MOS\,1 & 98 (91) [92] & 0.82 & 1.11 \\
(rev2663)  & MOS\,2 & 98 (92) [92] & 0.89 & 1.13 \\
\midrule
2014-06-25 & pn & 54 (54) [51] & 5.01 & 1.36 \\
0745110501 & MOS\,1 & 56 (56) [55] & 0.93 & 1.36 \\
(rev2664)  & MOS\,2 & 56 (56) [55] & 1.16 & 1.44 \\
\midrule
2014-06-29 & pn & 92 (92) [85] & 4.80 & 1.26 \\
0745110601 & MOS\,1 & 94 (94) [91] & 0.98 & 1.31 \\
(rev2666)  & MOS\,2 & 94 (94) [91] & 1.09 & 1.33 \\
\midrule
2014-07-07 & pn & 95 (95) [89] & 3.73 & 1.02 \\
0745110701 & MOS\,1 & 97 (97) [94] & 0.77 & 1.05 \\
(rev2670)  & MOS\,2 & 97 (97) [94] & 0.85 & 1.08 \\
\bottomrule
\end{tabular}
\caption{An observation log showing the broad-band durations / exposure times (ks), count rates (ct\,s$^{-1}$) and observed fluxes ($\times 10^{-11}$\,erg\,cm$^{-2}$\,s$^{-1}$) for the {\it XMM-Newton} observations of PG\,1211+143 detailed in this paper.  The filtered durations refer to the duration of each observation having filtered out background flares while the net exposures refer to the integrated exposure time having accounted for detector `dead time'.  All count rates and fluxes for the EPIC cameras are quoted over 0.2--10\,keV and are derived from the filtered datasets (where the count rate $=$ total counts / filtered duration).}
\label{tab:observation_log}
\end{table}

\subsubsection{EPIC-pn}

The bulk of our analysis focuses on data acquired with the European Photon Imaging Cameras (EPIC): the pn and the two Metal-Oxide Semiconductor (MOS) detectors.  These were operated in large-window mode ($\sim$94.9 per cent `livetime' for the pn; $\sim$99.5 per cent for the MOS) using the medium filter.  The background contribution was minimized by extracting source events from 20\,arcsec circular regions centred on the source\footnote{Source counts from 40\,arcsec circular regions were also extracted but the results were not found to significantly differ from those obtained with smaller regions, albeit with an additional $\sim$13\,per cent increase in count rate and a 400\,per cent increase in background counts in the source region.} while background events were extracted from much larger rectangular regions away from both the central source and other nearby background sources.  The background level was relatively stable for the majority of the observations although some degree of flaring was observed towards the end of rev2659, throughout rev2661 and at the beginning of rev2663 (see Fig.~\ref{fig:light_curve}).  To optimise data quality, events during these periods were initially filtered out by creating appropriate good time intervals.  This resulted in a total of $\sim$570\,ks of useful EPIC data.  Single and double good pixel events ({\tt PATTERN} $\leq$ 4) were utilized in our pn analysis while triple and quadruple pixel events ({\tt PATTERN} $\leq$ 12) were additionally used in our MOS analysis.  Response matrix files (RMFs) and ancillary response files (ARFs) were generated using the \textsc{sas} tasks \textsc{rmfgen} v2.2.1 and \textsc{arfgen} v1.90.4, respectively.  

The 0.2--10\,keV count rate averaged over all 7 orbits (having filtered out background flares) was 3.86\,ct\,s$^{-1}$ for the EPIC-pn and 0.80 and 0.89\,ct\,s$^{-1}$ for the EPIC MOS\,1 and 2 cameras, respectively, resulting in a total of $\sim$3.3$\times 10^{6}$ EPIC counts\footnote{We note that the background count rate obtained over the total energy band is $<$1 per cent of the source rate for all three EPIC cameras for each observation.}.  The broad-band observed flux (averaged over the three EPIC cameras) was $F_{\rm 0.2-10} = 1.08 \times 10^{-11}$\,erg\,cm$^{-2}$\,s$^{-1}$ ($F_{\rm 2-10} = 3.71 \times 10^{-12}$\,erg\,cm$^{-2}$\,s$^{-1}$), corresponding to an observed luminosity, $L_{\rm 0.2-10} = 1.8 \times 10^{44}$\,erg\,s$^{-1}$ ($L_{\rm 2-10} = 6.2 \times 10^{43}$\,erg\,s$^{-1}$).  The observed count rates, net exposure times and broad-band fluxes obtained from the individual EPIC observations are provided in Table~\ref{tab:observation_log}.

\subsubsection{OM}

{\it XMM-Newton} also features the co-aligned Optical Monitor (OM; \citealt{Mason01}), capable of providing simultaneous optical/UV coverage.  The OM uses of six primary filters: V (effective wavelength $=$ 5\,430\,\AA), B (4\,500\,\AA), U (3\,440\,\AA), UVW1 (2\,910\,\AA), UVM2 (2\,310\,\AA) and UVW2 (2\,120\,\AA).  For each observation we took a series of $\sim$2\,ks OM exposures in ``imaging'' mode (with a $2 \times 2$ pixel binning covering a 7' $\times$ 7' field of view), cycling through all six filters such that we acquired 1 exposure for each of the V, B, U, UVM2 and UVW2 filters while using the rest of the observation time with the UVW1 filter for simultaneous UV--X-ray monitoring.  Most observations resulted in typically $\sim$40 separate OM exposures, acquiring a total of 262 OM images with a total accumulated exposure of $\sim$523.8\,ks across all 7 observations.  An observation log is provided in Table~\ref{tab:om_log}.

\begin{table}
\centering
\begin{tabular}{l c c c}
\toprule
\multirow{3}{*}{Date} & \multirow{2}{*}{{\it XMM}} & Total OM & \multirow{2}{*}{Number of} \\
& \multirow{2}{*}{Revolution} & Exposure & \multirow{2}{*}{OM Images} \\
& & (ks) & \\
\midrule
2014-06-02 & rev2652 & 72.0 & 36 \\
2014-06-15 & rev2659 & 87.3 & 44 \\
2014-06-19 & rev2661 & 80.0 & 40 \\
2014-06-23 & rev2663 & 83.9 & 42 \\
2014-06-25 & rev2664 & 47.4 & 24 \\
2014-06-29 & rev2666 & 79.8 & 40 \\
2014-07-07 & rev2670 & 73.4 & 36 \\
\bottomrule
\end{tabular}
\caption{An observation log showing the total exposure and number of OM images acquired for each {\it XMM-Newton} observation of PG\,1211+143.  For each observation, there is 1 OM exposure acquired with each of the V, B, U, UVM2 and UVW1 filters while the remaining images are aquired with the UVW1 filter.}
\label{tab:om_log}
\end{table}

Since the observations were made in ``imaging'' mode, we processed the data using the \textsc{omichain}\footnote{\url{http://xmm.esac.esa.int/sas/current/doc/omichain/}} routine, as part of {\it XMM-Newton}'s \textsc{sas}.  This routine performs all the necessary processing for each exposure, such as flat-fielding, before running a source detection algorithm and computing the positions of each source within the field-of-view.  Standard aperture photometry is then applied to obtain count rates for each detected source - these count rates are then corrected for coincidence losses and detector dead time, before a final calibrated source list is produced.  The results are discussed in Section~\ref{sec:optical_and_uv_monitoring}.

\subsubsection{{\it Swift}}

Finally, we also obtained a series of observations with {\it Swift} \citep{Gehrels04}, as part of a Target of Opportunity (ToO) programme.  In total, {\it Swift} made 41 observations (typically separated by $\sim$1\,day), each with a typical duration of $\sim$1.5\,ks.  These observations began on 2014-06-04 and ended on 2014-08-04 and, hence, overlap with the 7 {\it XMM-Newton} observations.  We utilized data from {\it Swift}'s X-ray Telescope (XRT; \citealt{Burrows05}) and co-aligned Ultra-Violet/Optical Telescope (UVOT; \citealt{Roming05}).  The XRT was operating in ``photon counting mode'' and has a 23' $\times$ 23' FOV, providing useful data from 0.2 to 10\,keV.  The UVOT provides simultaneous UV/optical coverage (1\,700--6\,500\,\AA) in a 17' $\times$ 17' field.  Each {\it Swift} snapshot typically acquired data from exposures with three UVOT filters: U (central wavelength $=3\,465$\,\AA; 39 images), UVW1 (2\,600\,\AA; 40 images) and UVW2 (1\,928\,\AA; 41 images).  We note that the XRT was in ``anomaly mode'' from 2014-06-04 - 2014-06-10\footnote{\url{http://swift.ac.uk/support/anomaly.php}} rendering XRT data acquired during those dates unusable.  This resulted in a total of 37 usable XRT observations with a total exposure of $\sim$47.1\,ks.  

A visual inspection of the UVOT images reveals that they are steady, with 117 total usable frames split across 41 observations.  Each exposure was typically a couple of hundred seconds.  Source counts were extracted from PG\,1211+143 using the HEA{\sc soft}\footnote{\url{http://heasarc.nasa.gov/lheasoft/}} (v.6.16) task \textsc{uvotsource}, which performs aperture photometry and automatically performs necessary corrections such as coincidence loss and scaled background subtraction using the latest calibration database (CALDB).  We used a 5\,arcsec source radius and a 25\,arcsec background radius from a blank region of sky away from the source.  The {\it Swift} data are discussed in Section~\ref{sec:swift_monitoring}.

\section{Results} \label{sec:results}

\begin{figure*}
\begin{center}
\rotatebox{0}{\includegraphics[width=17.8cm]{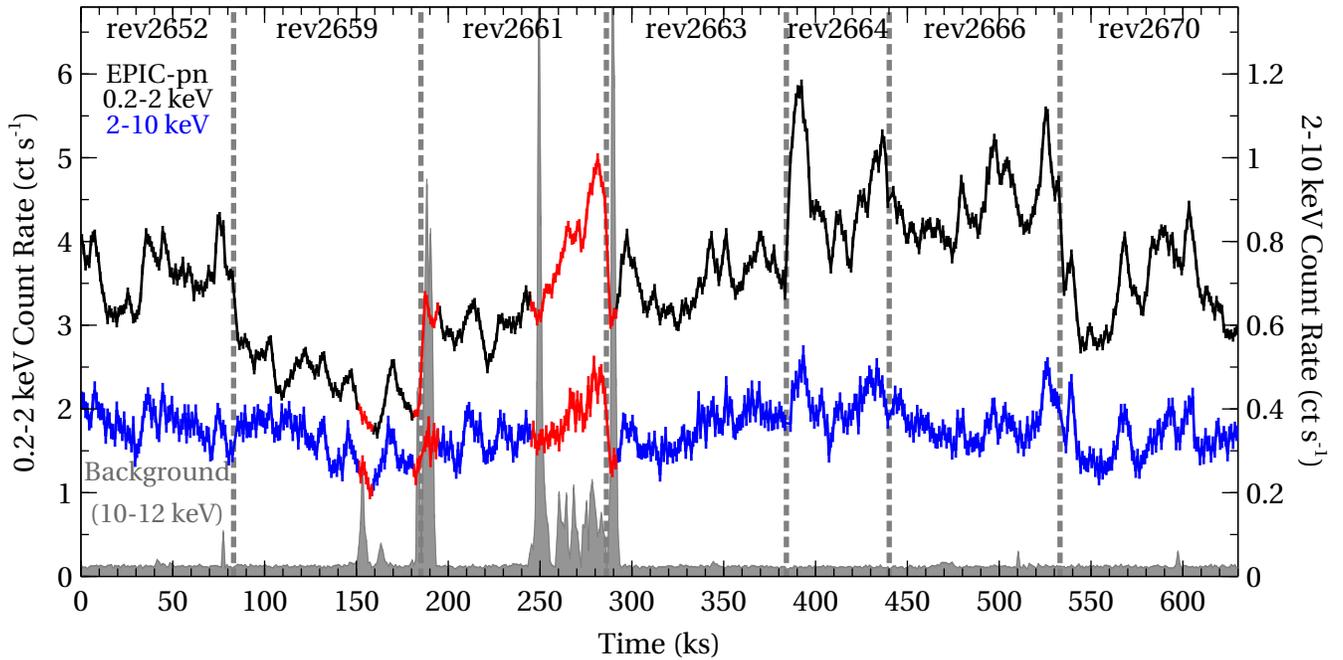}}
\end{center}
\vspace{-15pt}
\caption{The broad-band EPIC-pn lightcurve of PG\,1211+143 from all 7 observations made in 2014, shown with 1\,ks bins.  The black and blue curves show the 0.2--2 and 2--10\,keV lightcurves, respectively.  The red data points show the regions that were removed from the subsequent analysis due to significant background flaring while the grey curve shows the 10--12\,keV EPIC-pn background lightcurve acquired from the full FOV.  The vertical dotted lines show the beginning and end of each observation - note that these were not contiguous with a typical gap of a few days between observations.}
\label{fig:light_curve}
\end{figure*}

In this section we present the results of a detailed timing analysis of PG\,1211+143.  Custom \textsc{idl}\footnote{\url{http://www.star.le.ac.uk/sav2/idl.html}} scripts were used to extract light curves from individual EPIC event files.  These were background-subtracted and corrected for exposure losses while also interpolating over short telemetry drop outs where necessary. The EPIC light curve for PG\,1211+143 is shown in Fig.~\ref{fig:light_curve} with 1\,ks time bins in two energy bands: 0.2--2 and 2--10\,keV. The count rate is observed to vary by up to a factor of $\sim$3 over the entire 36-day period, while varying by up to $\sim$50\,per cent within individual observations (on $\sim$10\,ks time-scales).  The softer band is observed to be brighter and more highly variable, although the general pattern of variability on long timescales looks similar between the two bands.

\begin{figure}
\begin{center}
\rotatebox{0}{\includegraphics[width=8.4cm]{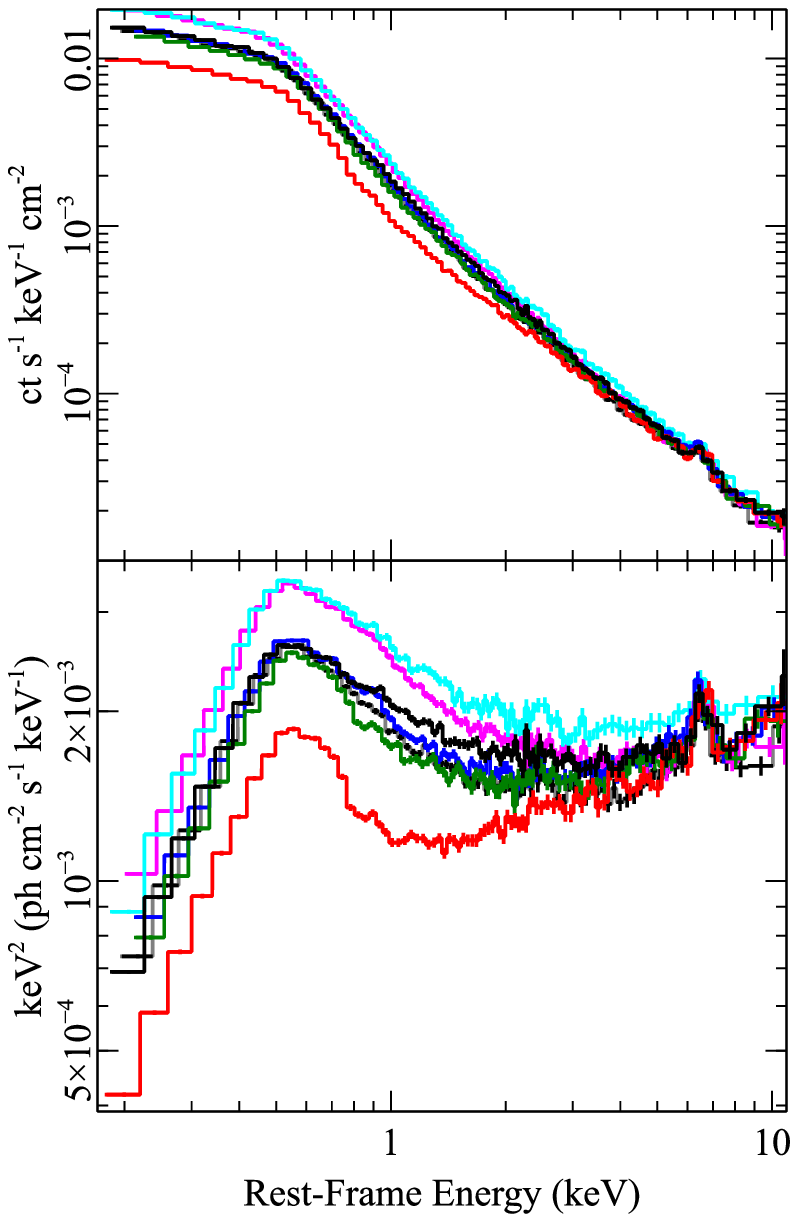}}
\end{center}
\vspace{-15pt}
\caption{The 7 time-averaged EPIC-pn spectra of PG\,1211+143 obtained in 2014: rev2652 (black), rev2659 (red), rev2661 (green), rev2663 (blue), rev2664 (cyan), rev2666 (magenta) and rev2670 (grey).  In the lower panel, the spectra are ``fluxed'' and are binned up coarsely for clarity.}
\label{fig:2014_spectra}
\end{figure}

A detailed investigation into the spectral properties is undertaken in \citet{Pounds15a} and \citet{Pounds15b} and additional forthcoming papers - however, we discuss the broad spectral shape of PG\,1211+143 here for comparison with the timing properties.  Fig.~\ref{fig:2014_spectra} shows all 7 EPIC-pn spectra obtained in 2014, having divided out the effective area.  The spectra were binned using \textsc{specgroup} to match the detector resolution (using bins of approximately $1/3$ of the FWHM) with the additional criterion that each bin contain at least 25 source counts in order to make use of standard $\chi^{2}$ fitting methods.  The observation-averaged source flux varies by $\sim$50\,per cent between observations with the bulk of the long-term variability apparently taking place $< 2$\,keV.  

To first order, the general shape of the underlying continuum can be decomposed into three primary spectral components: (i) an absorbed intrinsic power law with photon index, $\Gamma \sim 1.6$, which dominates the hard X-ray spectrum; (ii) a component of soft excess to account for the upturn in the soft X-ray continuum (e.g. a blackbody component with temperature, $k{\rm T} \sim 0.1$\,keV), and (iii) an unabsorbed soft power-law component of $\Gamma \sim 2.9$, which becomes significant at $\lesssim 2$\,keV and dominates the soft X-ray spectrum.  Superimposed on this broad-band continuum is then an array of discrete features arising from the warm absorber / high-velocity outflow and the Fe\,K emission profile at $\sim$6--7\,keV (see \citealt{Pounds15a} and additional forthcoming papers for further details). 

\begin{figure}
\begin{center}
\rotatebox{0}{\includegraphics[width=8.4cm]{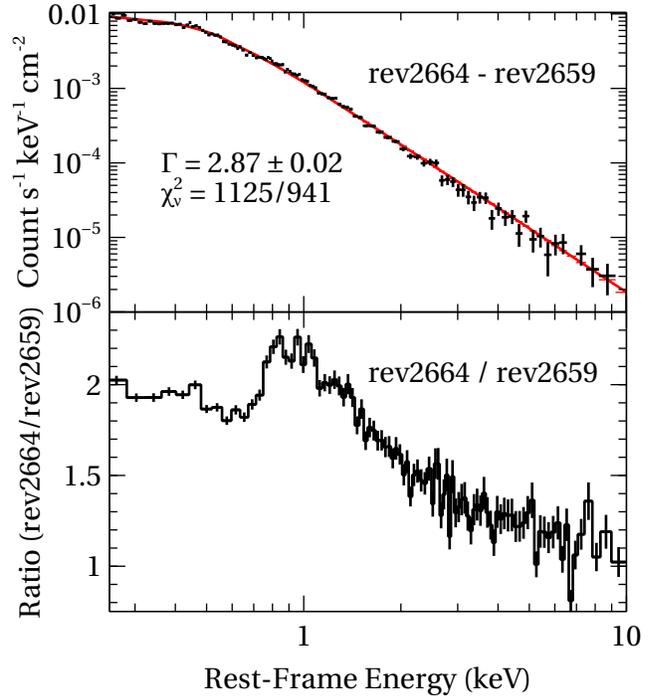}}
\end{center}
\vspace{-15pt}
\caption{Upper panel: the difference spectrum of PG\,1211+143 having subtracted the lowest flux observation (rev2659) from the highest flux observation (rev2664).  The resultant spectrum is fitted with a steep power law (modified by Galactic absorption) with a photon index of $\Gamma = 2.87 \pm 0.02$.  Lower panel: rev2664/rev2659 ratio spectrum.  Both spectra have been binned up for clarity.}
\label{fig:difference_ratio_spectra}
\end{figure}

The general spectral shape between orbits can largely be accounted for by allowing for different relative normalizations of the three components.  In particular, the long-term spectral variability  (i.e. between observations on timescales $> 1$\,day) is dominated by changes in the normalization of the soft power-law component.  This is illustrated in the upper panel of Fig.~\ref{fig:difference_ratio_spectra}, which shows the EPIC-pn `difference spectrum' of PG\,1211+143.  Here, the lowest flux observation (rev2659) has been subtracted from the highest flux observation (rev2664).  The resultant spectrum takes the form of a steep continuum which can be reasonably well fitted by a power law of photon index $\Gamma = 2.87 \pm 0.02$ ($\chi^{2}_{\nu} = 1125/941$).  This behavior is largely consistent when measuring the difference spectrum between any two observations.

\begin{figure}
\begin{center}
\rotatebox{0}{\includegraphics[width=8.4cm]{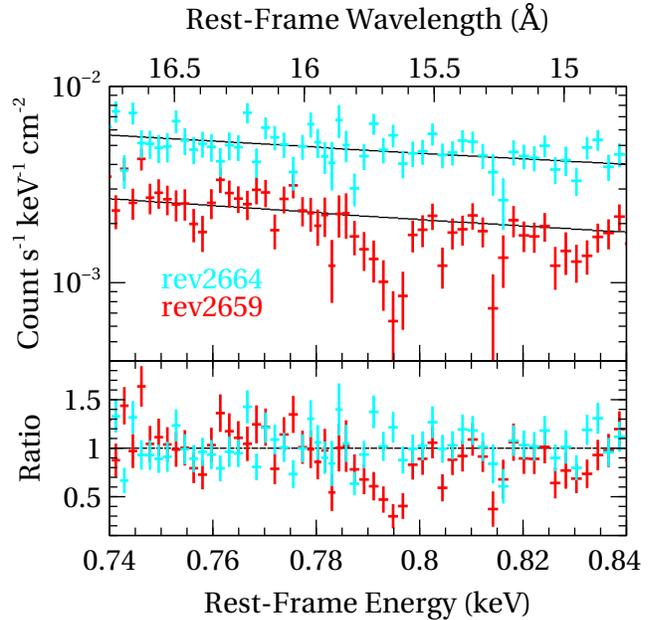}}
\end{center}
\vspace{-15pt}
\caption{The summed RGS\,1+2 spectra of PG\,1211+143 from revs 2659 (red) and 2664 (cyan) over the 0.74--0.84\,keV energy range.  The spectra have been locally fitted with a simple power-law component to highlight the variable absorption at $\sim$0.79\,keV, which can be observed in the ratio of the residuals to the model (lower panel).  The data are binned to approximately $1/3$ of the FWHM of the detector resolution.}
\label{fig:variable_absorption}
\end{figure}

In addition to the difference spectrum, the lower panel of Fig.~\ref{fig:difference_ratio_spectra} shows the ratio spectrum.  Here, the rev2664 EPIC-pn spectrum has been divided by the rev2659 spectrum.  The ratio between the two spectra is close to unity at high energies, consistent with the hard power-law component remaining (on average) steady in flux over time.  At lower energies, however, the spectra diverge as the soft component dominates the variability.  Curiously, we note the presence of some structure in the ratio spectrum at $\gtrsim$0.7\,keV, which takes the form of a sharp rise going upwards in energy.  This feature likely arises from a component of strongly variable ionized absorption in the X-ray spectrum of PG\,1211+143.  This is shown in Fig.~\ref{fig:variable_absorption} and is perhaps dominated by the Fe\,\textsc{i-xvi} unresolved transition array (UTA), which appears to be significantly variable in the observed spectrum on timescales of days and is strongest when the source flux is weakest.  Detailed modelling of these spectral features along with the associated spectral variability is provided in a number of companion papers (e.g. \citealt{Pounds15a, Pounds15b, Lobban16}).  Finally, we also note that the broad-band optical--UV--X-ray spectral energy distribution (SED) is shown and roughly parametrized in Section~\ref{sec:sed}.  However, we now proceed to investigate the timing properties.

\subsection{The power spectrum} \label{sec:psd}

\begin{figure}
\begin{center}
\rotatebox{0}{\includegraphics[width=8.4cm]{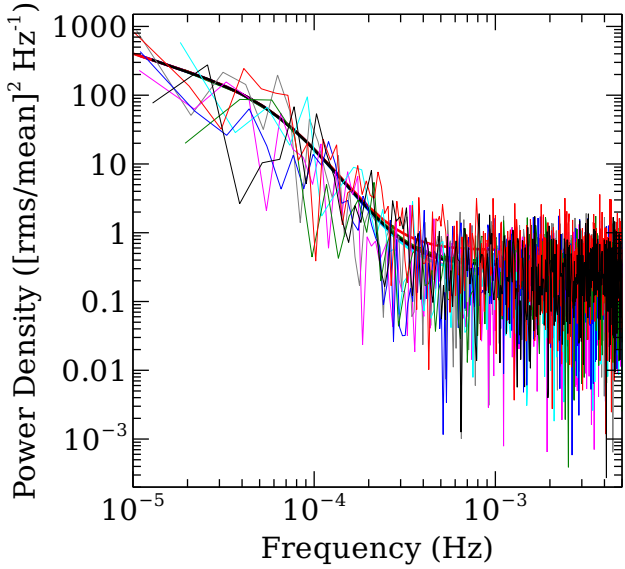}}
\end{center}
\vspace{-15pt}
\caption{The 0.2--10\,keV PSD of PG\,1211+143 in the form of ``raw'' unbinned periodograms from all 7 EPIC pn+MOS observations: rev2652 (black), rev2659 (red), rev2661 (green), rev2663 (blue), rev2664 (cyan), rev2666 (magenta) and rev2670 (grey).  The solid curves show the fitted bending power-law model described in Section~\ref{sec:psd}.}
\label{fig:psd}
\end{figure}

We firstly investigated the PSD, which provides an estimate of the variability power as a function of temporal frequency.  We extracted background-subtracted light curves for all 7 {\it XMM-Newton} observations with a timing resolution of $\Delta t = 100$\,s, utilizing data from all 3 EPIC cameras (pn + MOS).  We then computed simple periodograms (\citealt{Priestly81}; \citealt{PercivalWalden93}; \citealt{Vaughan03}) from each orbit in (rms/mean)$^{2}$ units (see Fig.~\ref{fig:psd}).  The long duration of some observations (i.e. $\sim$100\,ks) allows us to estimate the PSD down to frequencies as low as $\sim$10$^{-5}$\,Hz, which is approximately a factor of 2 lower than possible with previous archival data for this source (e.g. \citealt{Gonzalez-MartinVaughan12}).

We simultaneously fitted all 7 ``raw'' unbinned periodograms within \textsc{xspec}.  For a given model, we found the best-fitting model parameters by minimizing the Whittle likelihood, $S$, as discussed in \citet{Vaughan10} and \citet{BarretVaughan12}.  We fitted the broad-band 0.2--10\,keV PSD with two models: (i) a simple power law, and (ii) a bending power law (see \citealt{Gonzalez-MartinVaughan12} for discussion of the models and fitting methods).  

Model (i) has three free parameters: the spectral index ($\alpha$), the power-law normalization and an additive constant of zero slope to account for the Poisson noise at high frequencies.  We tied all parameters together between observations except for the normalization of the constant component to allow for varying levels of Poisson noise due to differing count rates between orbits.  The single power-law model results in a Whittle fit statistic of $S = -62285.6$ for 2\,788 degrees of freedom (d.o.f.) and has a best-fitting slope of $\alpha = 2.52 \pm 0.10$, dominating at frequencies $\nu \lesssim 4 \times 10^{-4}$\,Hz.

We then allowed for the presence of a bend in the power law by fitting model (ii), which consists of five parameters: the normalization, the low-frequency spectral index (below the bend; $\alpha_{\rm 1}$), the high-frequency slope ($\alpha_{\rm 2}$), the bend frequency ($\nu_{\rm b}$) and, again, a constant to account for Poisson noise.  These {\it XMM-Newton} data are somewhat insensitive to the low-frequency slope - however, we find it to be constrained to $\alpha_{\rm 1} = 0.9^{+0.3}_{-0.1}$, which is consistent with the value of 1 typically found from long-term monitoring campaigns (e.g. \citealt{UttleyMcHardyPapadakis02}, \citealt{Markowitz03}, \citealt{McHardy06}).  The high-frequency slope has a best-fitting value of $\alpha_{\rm 2} = 3.5^{+0.4}_{-0.3}$, consistent with the average value of 3.08 obtained from a sample of 15 Seyfert galaxies by \citet{Gonzalez-MartinVaughan12}.  The Poisson noise then begins to dominate at frequencies $\gtrsim 4 \times 10^{-4}$\,Hz.  The bend occurs at a frequency of $\nu_{\rm b} = 6.9^{+2.4}_{-1.9} \times 10^{-5}$\,Hz and the inclusion of this feature improves the fit statistic by $\Delta S = 37.6$ when compared to the single power-law model.  Since $\Delta S$ behaves approximately as $\Delta \chi^{2}$, this suggests that the bend is detected with high significance.  This fit to the PSD is shown in Fig.~\ref{fig:psd} and the best-fitting model parameters are provided in Table~\ref{tab:psd}.  Errors are quoted at the 1$\sigma$ level (i.e. $\Delta S = 1.0$ for one interesting parameter).

We then fitted the above models to PSDs generated in `soft' (0.2--2\,keV) and `hard' (2--10\,keV) energy bands but found no significant energy dependence of the PSD with best-fitting values typically consistent within the uncertainties.  The best-fitting values are shown in Table~\ref{tab:psd}.  Additionally, we tried fitting the 0.2--2 and 2--10\,keV PSDs simultaneously, tying the bend frequency and low-frequency slope between energy bands, while allowing the high-frequency slope to vary.  We find that the high-frequency slope is consistent within the errors between the two energy bands: $\alpha^{\rm 0.2-2}_{2} = 3.7^{+0.4}_{-0.3}$ versus $\alpha^{\rm 2-10}_{2} = 3.2^{+0.5}_{-0.3}$.  Meanwhile, we find best-fitting values for the low-frequency slope and bend frequency in this simultaneous fit of $\alpha_{\rm 1} = 0.8^{+0.3}_{-0.2}$ and, $\nu_{\rm b} = 7.0^{+1.9}_{-1.6} \times 10^{-5}$\,Hz, respectively.  Allowing parameters ($\alpha_{\rm 1}$, $\nu_{\rm b}$, $\alpha_{\rm 2}$) to vary independently between datasets, we are unable to detect any significant differences between orbits. 

\begin{table}
\centering
\begin{tabular}{l c c c c}
\toprule
Model & \multirow{2}{*}{Parameter} & \multicolumn{3}{c}{Energy Range (keV)} \\
(Statistic) & & 0.2--10 & 0.2--2 & 2--10 \\
\midrule
Power Law & $\alpha$ & $2.5^{+0.1}_{-0.1}$ & $2.5^{+0.1}_{-0.1}$ & $2.1^{+0.1}_{-0.1}$ \\
(Whittle) & $S$ & -62285.6 & -61639.3 & -50244.9 \\
\midrule
\multirow{2}{*}{Bending} & $\alpha^{\rm low}_{\rm 1}$ & $0.9^{+0.3}_{-0.1}$ & $0.8^{+0.4}_{-0.1}$ & $1.1^{+0.3}_{-0.3}$ \\
\multirow{2}{*}{Power Law} & $\nu_{\rm b}$ ($\times 10^{-5}$\,Hz) & $6.9^{+2.4}_{-1.9}$ & $6.5^{+2.6}_{-1.9}$ & $11.2^{+3.2}_{-3.1}$ \\
& $\alpha^{\rm high}_{\rm 2}$ & $3.5^{+0.4}_{-0.3}$ & $3.5^{+0.4}_{-0.3}$ & $4.5^{+1.3}_{-0.9}$ \\
\multirow{2}{*}{(Whittle)} & $S$ & -62323.2 & -61674.0 & -50273.0 \\
& $\Delta S$ & 37.6 & 34.7 & 29.0 \\
\bottomrule
\end{tabular}
\caption{The best-fitting parameters of the simple power-law and bending power-law models fitted to the PSD of PG\,1211+143 in the 0.2--10, 0.2--2 and 2--10\,keV energy ranges.  The Whittle statistic, $S$, is quoted for each model, while $\Delta S$ refers to the difference between the two models; i.e. the improvement in the fit with the inclusion of the bend.  See Section~\ref{sec:psd} for details.}
\label{tab:psd}
\end{table}

\subsection{The rms-flux relation} \label{sec:rms-flux}

A common feature of the observed X-ray variability in XRBs, AGN, ultraluminous X-ray sources (ULXs) and cataclysmic variables (CVs) is the `rms-flux' relation.  This is a relationship which shows that the absolute root-mean-square (rms) amplitude of variability scales linearly with the X-ray flux of the source, such that sources display more variability when they are brighter (e.g. \citealt{UttleyMcHardy01,Gleissner04,HeilVaughan10}).  Here, we investigate the rms-flux behaviour for PG\,1211+143.

\begin{figure}
\begin{center}
\rotatebox{0}{\includegraphics[width=8.4cm]{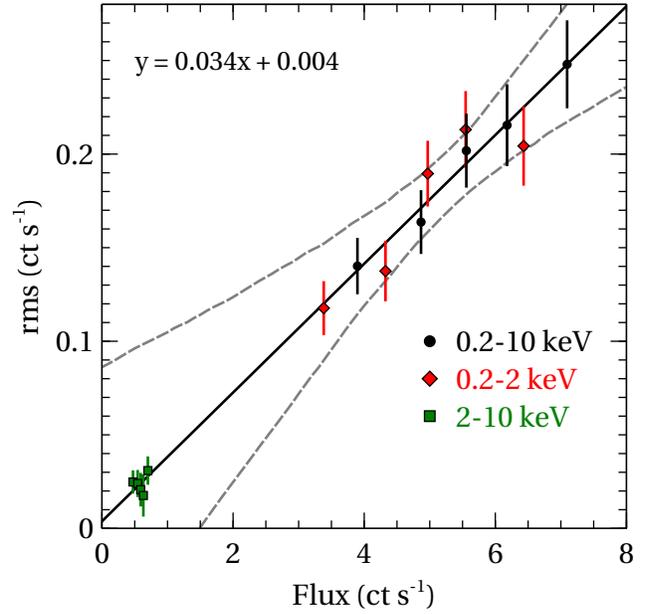}}
\end{center}
\vspace{-15pt}
\caption{The rms-flux relation of PG\,1211+143 across three energy bands: 0.2--10\,keV (black circles), 0.2--2\,keV (red diamonds) and 2--10\,keV (green squares).  The black solid line is a line of best-fit fitted to the 0.2--10\,keV data ($y = Ax + B$, where $A = 0.034 \pm 0.008$ and $B = 0.004 \pm 0.041$).  The grey dashed lines enclose the 95\,per cent confidence interval.  See Section~\ref{sec:rms-flux} for details.}
\label{fig:rms-flux}
\end{figure}

We use the EPIC-pn and EPIC MOS data, binned with a timing resolution, $dt = 50$\,s and split into segments, 5\,ks in length.  For each segment, we firstly compute the periodogram and the mean count rate.  We then calculate the PSD as a function of flux by averaging together the periodograms over 5 count rate bins.  We then compute the rms in each bin by taking the square root, having integrated under the average PSD and subtracted the Poisson noise.  The rms-flux relation for PG\,1211+143 is shown in Fig.~\ref{fig:rms-flux} across three energy bands: 0.2--10 (broad-band), 0.2--2 and 2--10\,keV.

We tested the linearity of the relationship by fitting a straight line to the data across the broad-band 0.2--10\,keV energy range.  This appears to fit the data well, with a best-fitting slope of $0.034 \pm 0.008$ (corresponding to 3.4\,per cent rms on timescales of 0.1--5\,ks) and a y-intercept of $0.004 \pm 0.041$, consistent with zero.  The 95\,per cent uncertainty on the linear model is illustrated by the grey dashed lines.  These were estimated from the 97.5 and 2.5\,per cent rms values for a given flux value from 500 models randomly generated from the parameter distribution (i.e. using the best-fitting values and the covariance matrix).  While the positive rms-flux relation is well-detected, it is dominated by the brighter, soft band (0.2--2\,keV).  The relationship is less clear in the fainter (and, accordingly, less variable) hard band, although still consistent with the best-fitting slope.  We did also investigate the rms-flux behaviour within individual orbits.  While the positive relation is still largely recovered, the smaller range in flux and far fewer number of segments with which to average over per orbit make the relationship difficult to constrain.  Likewise, we largely recover the positive correlation by measuring the rms-flux relation on long timescales by plotting the rms against the flux obtained from each individual observation.

\subsection{The rms spectrum} \label{sec:rms-spectrum}

\begin{figure}
\begin{center}
\rotatebox{0}{\includegraphics[width=8.4cm]{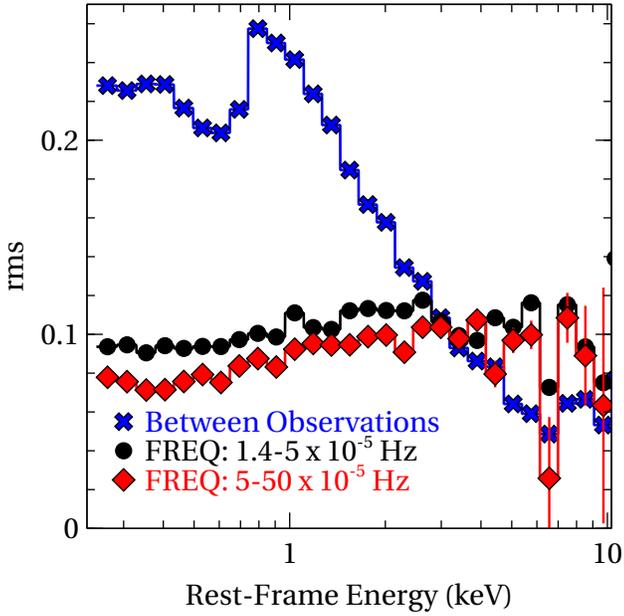}}
\end{center}
\vspace{-15pt}
\caption{The EPIC-pn+MOS 0.2--10\,keV fractional rms spectrum of PG\,1211+143 averaged over all orbits.  The black circles show the fractional rms spectrum over the $\sim$1.4--5 $\times 10^{-5}$\,Hz frequency range and the red diamonds show the higher 5--50 $\times 10^{-5}$\,Hz frequency range.  Meanwhile, the blue crosses show the long-term `between-observation' fractional rms spectrum.  See Section~\ref{sec:rms-spectrum} for details.}
\label{fig:rms-spectrum}
\end{figure}

Here we show the rms variability amplitude as a function of energy (i.e. the `rms spectrum') using the combined EPIC-pn+MOS data averaged over all orbits (see \citealt{Vaughan03} and references therein).  The rms spectrum was computed by extracting light curves in 28 energy bands approximately equally spaced in log$(E)$ over the 0.2--10\,keV energy range using equal-length light curve segments.  For each energy band and segment, we then calculated the fractional excess variance ($\sigma_{\rm xs}^2$ / mean$^{2}$) and averaged across each segment.  The fractional rms was then calculated by taking the square root of the excess variance and is shown in Fig.~\ref{fig:rms-spectrum} over two separate frequency bands: $\sim$1.4--5 $\times 10^{-5}$\,Hz ($\Delta t = 10$\,ks; 70\,ks segments) and 5--50 $\times 10^{-5}$\,Hz ($\Delta t = 1$\,ks; 20\,ks segments).  The fractional rms spectrum has a similar shape across both high and low frequencies in that it is approximately constant $> 2$\,keV while the fractional rms falls slightly at lower energies.  However, the amplitude of the rms spectrum is higher over the lower frequency band.  We note the sharp drop in variability at $\sim$6--7\,keV which is most likely due to a constant component of emission from Fe\,K which is not responding to the continuum on the timescales probed here.  We also show the fractional long-term `between-observations' rms spectrum in order to quantify the variability on much longer timescales.  This provides us with an alternative way to view the clear spectral variations seen in Fig.~\ref{fig:2014_spectra} with a characteristic ``softer when brighter'' pattern with much smaller variations in the harder bands.  As Fig.~\ref{fig:rms-spectrum} shows, on these timescales of $\gtrsim$\,days, the variability is much stronger at low energies, peaking at $\sim$0.8\,keV, while dropping off at higher energies.  Again, the fractional rms is lowest at $\sim$6--7\,keV where the Fe\,K emission complex dominates. 

\subsection{Optical and UV monitoring} \label{sec:optical_and_uv_monitoring}

We begin by discussing data acquired with the OM on-board {\it XMM-Newton}.  As stated in Section~\ref{sec:observations_and_data_reduction}, we obtained one 2\,ks exposure with each of the V, B, U, UVM2 and UVW2 filters per observation while we used the rest of the observing time for UV monitoring with the UVW1 filter.  This resulted in a total of 227 OM images with the UVW1 filter.  We begin by investigating the within-observation variability across all 7 {\it XMM-Newton} observations.  We note that there will be a significant contribution by the host galaxy to the observed nuclear light - however, this should remain largely constant. 

\begin{figure*}
\begin{center}
\rotatebox{0}{\includegraphics[width=17.8cm]{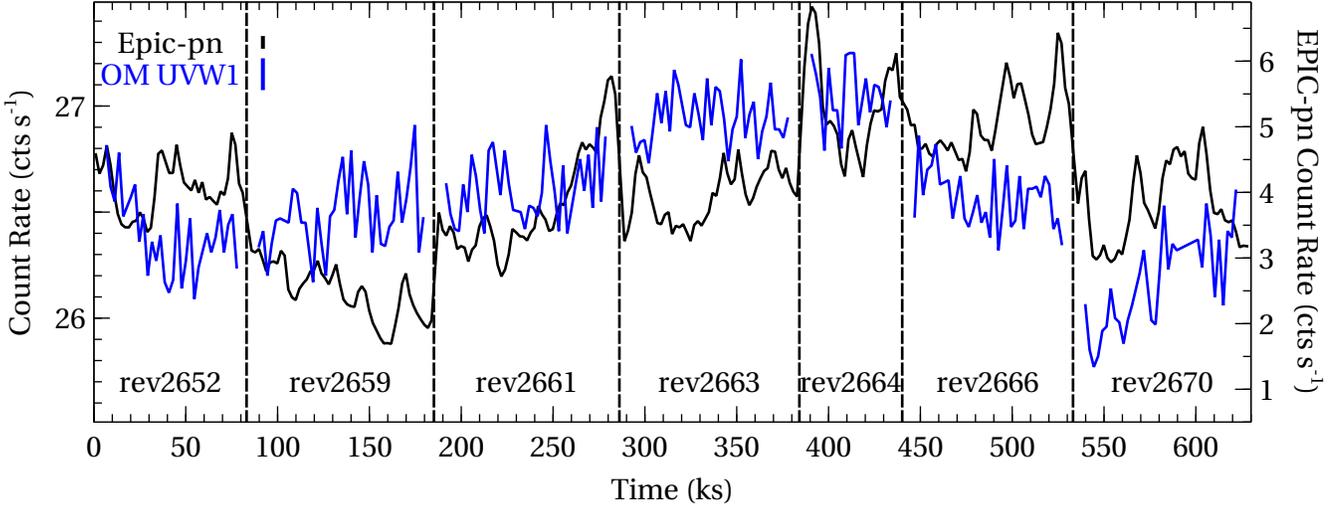}}
\end{center}
\vspace{-15pt}
\caption{The concatenated OM light curve of PG\,1211+143, consisting of all 7 observations acquired in 2014.  The OM data are acquired from monitoring with the UVW1 filter and are shown in blue with a timing resolution of $\Delta t = 2$\,ks.  The EPIC-pn data light curve (X-ray: 0.2--10\,keV) is overlaid in black for comparison ($\Delta t = 2$\,ks).  The dashes in towards the top-left corner are representative error bars.}
\label{fig:uvw1_pn_lc}
\end{figure*} 

The UV light curve of PG\,1211+143 is shown in Fig.~\ref{fig:uvw1_pn_lc}.  As no additional timing information is available within each individual 2\,ks exposure, we use the minimum possible time bin size: $\Delta t = 2$\,ks.  Variability in the UV flux is observed over this 5-week period which can be roughly characterised as a gradual increase in flux over the first five observations, peaking in rev2664, before tailing off in the final two observations.  Weaker, short-term (`within-observation') variability is superimposed over this trend\footnote{We note that the typical uncertainty in each time bin is $\sim$$\pm 0.15$\,ct\,s$^{-1}$}.

In the case of the UVW1 variability in PG\,1211+143, the fractional variability (e.g. \citealt{Vaughan03}) is $F_{\rm var} \sim 1$\,per cent, while the 0.2--10\,keV X-ray variability is much stronger with $F_{\rm var} \sim 21.6$\,per cent.  As such, the relative long-term X-ray variability is roughly 20 times greater than the observed UV variability.  The total EPIC-pn X-ray light curve is shown overlaid on the UV light curve in Fig.~\ref{fig:uvw1_pn_lc}. 

\begin{figure}
\begin{center}
\rotatebox{0}{\includegraphics[width=8.4cm]{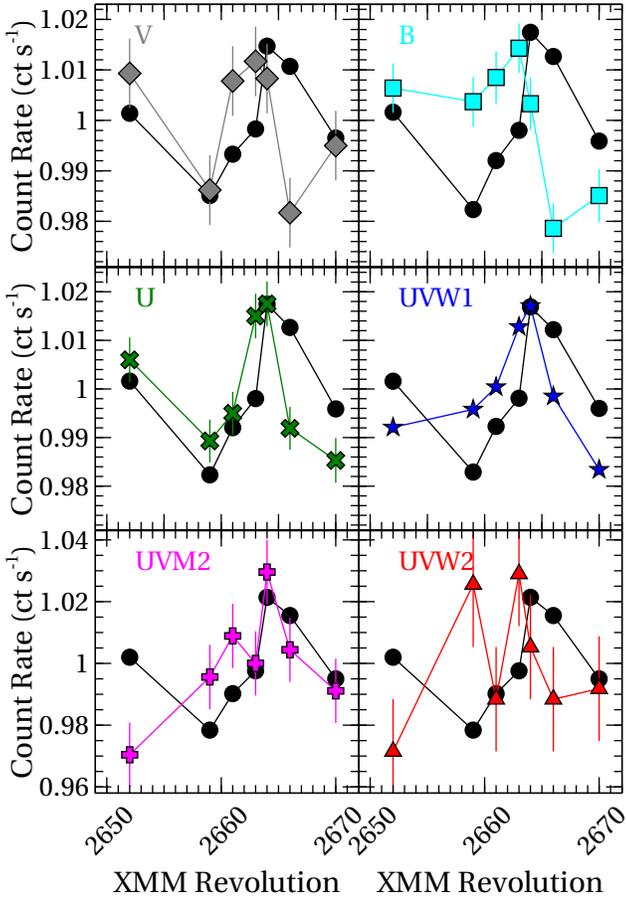}}
\end{center}
\vspace{-15pt}
\caption{The optical/UV count rates for each of the six OM filters versus the 0.2--10\,keV EPIC-pn count rates for each observation.  In each case the OM and X-ray count rates have been normalized to their own mean values, while the amplitude of X-ray variability has been reduced such that $F_{\rm var}$ is identical for both datasets being compared.}
\label{fig:om_x-ray_normalised}
\end{figure}

We also investigated the long-term ``between-observation'' optical / UV variability using all six OM filters.  For each of the V, B, U, UVM2 and UVW2 filters, we only obtained one 2\,ks exposure per observation, where the observations are typically separated by a few days and span a total $\sim$5-week period.  The corrected count rates are given in Table~\ref{tab:om_count_rates}.  Note that the UVW1 count rates are the mean values for each observation.  From the observed count rates, we can estimate the fractional variability, $F_{\rm var}$, which is typically $\sim$1--2\,per cent in each bandpass.  In Fig.~\ref{fig:om_x-ray_normalised}, we plot the count rates obtained with each filter versus the X-ray count rate, as a function of time.  In order to make a visual comparison, we normalized the count rates to their own mean values and reduced the amplitude of the X-ray variability such that the $F_{\rm var}$ values were identical.  We propagated the uncertainties accordingly in each case.  

We tested for any optical/UV/X-ray correlations between observations with the Pearson correlation coefficient, $r$, which is a measure of the strength of the linear relationship between two variables.  The correlation coefficients for the six OM filters versus the 0.2--10\,keV EPIC-pn count rates are: V $= 0.114$ ($p = 0.81$), B $= -0.352$ ($p = 0.44$), U $= 0.498$ ($p = 0.26$), UVW1 $= 0.433$ ($p = 0.33$), UVM2 $=0.415$ ($p = 0.35$) and UVW2 $= -0.296$ ($p = 0.52$).  As such, the optical/UV - X-ray fluxes do not appear to be significantly correlated, although we note that 7 snapshots over a five-week period may not be sufficient to detect a correlation, particularly given the strong short-term variability exhibited by the X-rays. 

Finally, we roughly estimated the observed flux in each bandpass using the count-rate-to-flux conversion factors provided by the \textsc{sas} team\footnote{\url{http://xmm.esac.esa.int/sas/9.0.0/watchout/Evergreen\_tips\_and\_tricks/uvflux.shtml}}.  These are calculated in units of $\times 10^{-15}$\,erg\,cm$^{-2}$\,s$^{-1}$\,\AA$^{-1}$ and are listed in Table~\ref{tab:om_count_rates}.

\begin{table*}
\centering
\begin{tabular}{l c c c c c c}
\toprule
{\it XMM} & \multicolumn{6}{c}{Corrected Count Rate per OM Filter (ct\,s$^{-1}$)} \\
Revolution & V (5\,430\,\AA) & B (4\,500\,\AA) & U (3\,440\,\AA) & UVW1 (2\,910\,\AA) & UVM2 (2\,310\,\AA) & UVW2 (2\,120\,\AA) \\
\midrule
rev2652 & $20.54 \pm 0.14$ & $44.90 \pm 0.22$ & $52.74 \pm 0.24$ & $26.39 \pm 0.03$ & $6.56 \pm 0.07$ & $2.87 \pm 0.05$ \\
rev2659 & $20.07 \pm 0.14$ & $44.78 \pm 0.22$ & $51.86 \pm 0.23$ & $26.49 \pm 0.02$ & $6.73 \pm 0.07$ & $3.03 \pm 0.06$ \\
rev2661 & $20.51 \pm 0.14$ & $44.99 \pm 0.23$ & $52.16 \pm 0.23$ & $26.61 \pm 0.02$ & $6.82 \pm 0.07$ & $2.92 \pm 0.05$ \\
rev2663 & $20.59 \pm 0.14$ & $45.25 \pm 0.22$ & $53.21 \pm 0.24$ & $26.94 \pm 0.02$ & $6.76 \pm 0.07$ & $3.04 \pm 0.05$ \\
rev2664 & $20.52 \pm 0.14$ & $44.76 \pm 0.23$ & $53.34 \pm 0.24$ & $27.05 \pm 0.03$ & $6.96 \pm 0.07$ & $2.97 \pm 0.05$ \\
rev2666 & $19.98 \pm 0.14$ & $43.66 \pm 0.22$ & $52.00 \pm 0.23$ & $26.56 \pm 0.03$ & $6.79 \pm 0.07$ & $2.92 \pm 0.05$ \\
rev2670 & $20.25 \pm 0.14$ & $43.95 \pm 0.23$ & $51.65 \pm 0.24$ & $26.16 \pm 0.03$ & $6.70 \pm 0.07$ & $2.93 \pm 0.05$ \\
\midrule
Mean (ct\,s$^{-1}$) & $20.35 \pm 0.06$ & $44.61 \pm 0.08$ & $52.42 \pm 0.09$ & $26.60 \pm 0.01$ & $6.76 \pm 0.03$ & $2.95 \pm 0.02$ \\
$F_{\rm var}$\,(per cent) & $\sim$1.0 & $\sim$1.2 & $\sim$1.2 & $\sim$1.0 & $\sim$1.5 & $\sim$1.2 \\
Flux ($\times 10^{-15}$\,erg\,cm$^{-2}$\,s$^{-1}$\,\AA$^{-1}$) & $5.07 \pm 0.01$ & $5.75 \pm 0.01$ & $10.17 \pm 0.02$ & $12.66 \pm 0.01$ & $14.87 \pm 0.07$ & $16.84 \pm 0.11$ \\
\bottomrule
\end{tabular}
\caption{The corrected count rates for each of the filters utilized by the OM for each observation.  The effective wavelengths of each filter are given in parentheses.  The UVW1 filter was used for monitoring - as such, the stated count rates are the mean rates for each observation.  The bottom three rows display the mean count rates, fractional variability, $F_{\rm var}$, and the flux estimated over each filter bandpass across all 7 observations.}
\label{tab:om_count_rates}
\end{table*}

\subsubsection{Optical--X-ray SED} \label{sec:sed}

\begin{figure}
\begin{center}
\rotatebox{0}{\includegraphics[width=8.4cm]{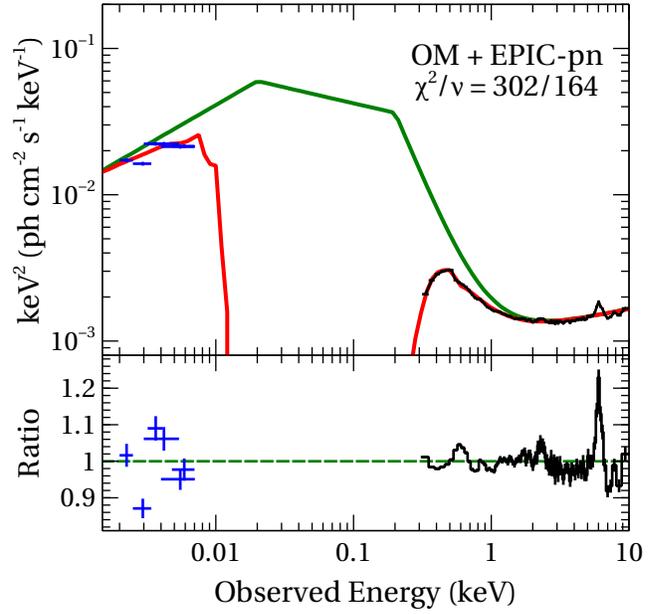}}
\end{center}
\vspace{-15pt}
\caption{Upper panel: the time-averaged OM (blue) + EPIC-pn (black) SED of PG\,1211+143 fitted with a model of the form: \textsc{redden} $\times$ \textsc{tbabs}$_{\rm Gal}$ $\times$ \textsc{tbabs} $\times$ (\textsc{bkn2pow} + \textsc{po}) (red).  Lower panel: the data/model ratio of the residuals to this model.  The solid green line in the upper panel shows the intrinsic unabsorbed, de-reddened SED model.  See Section~\ref{sec:sed} for details.}
\label{fig:sed}
\end{figure}

Here, we use the time-averaged OM data to show the broad-band optical--UV--X-ray SED of PG\,1211+143.  We accounted for the effective area of each filter by using the appropriate canned response matrices\footnote{\url{http://xmm2.esac.esa.int/external/xmm_sw_cal/calib/om_files.shtml}}.  The SED is shown in Fig.~\ref{fig:sed}.

By fitting within \textsc{xspec} v12.8.2 \citep{Arnaud96}, the rough shape of the time-averaged background-subtracted SED can be very loosely described by a model consisting of the form: \textsc{redden} $\times$ \textsc{tbabs}$_{\rm Gal}$ $\times$ \textsc{tbabs} $\times$ (\textsc{bkn2pow} + \textsc{po}).  \textsc{tbabs} models absorption by neutral gas and dust \citep{WilmsAllenMcCray00}, utilizing the photoionization absorption cross-sections of \citet{Verner96}.  The \textsc{tbabs}$_{\rm Gal}$ component accounts for the Galactic hydrogen column, which is fixed at a value of $3.06 \times 10^{20}$\,cm$^{-2}$ based on the measurements of \citet{Kalberla05} at the position of this source and modified to take in account the effect of molecular hydrogen (H$_{\rm 2}$), according to \citet{Willingale13}.  The secondary \textsc{tbabs} component accounts for any additional absorption from cold material beyond the Galactic column (found to be $N_{\rm H} \sim 3 \times 10^{20}$\,cm$^{-2}$).  The \textsc{redden} component accounts for IR (infrared) / optical / UV extinction \citep{CardelliClaytonMathis89} and was fixed at $E(B-V) = 0.035$.  Meanwhile, the \textsc{bkn2pow} represents a twice-broken power law while the \textsc{po} accounts for the hard X-ray continuum.  The \textsc{bkn2pow} component has a photon index of $\Gamma_{\rm 1} \sim 1.5$ to account for the optical--UV continuum before breaking first at 0.02\,keV ($\Gamma _{\rm 2} \sim 2.4$) and then again at 0.2\,keV into a steeper component ($\Gamma_{\rm 3} \sim 3.9$) to account for the soft X-ray excess.  Then, the \textsc{po} component accounts for the hard X-ray continuum ($\Gamma \sim 1.8$), dominating $\gtrsim 1$\,keV. 

While outlining the general shape of the broad-band SED, the fit is clearly poor, with $\chi^{2}/\nu = 302/164$ (having allowed a 3\,per cent systematic error, mainly for the OM data\footnote{The 3\,per cent systematic is larger than the variations seen in the OM/UVOT data.  However, this is to account for calibration errors in the inter-band absolute flux scale.}).  This is largely due to a considerable oversimplification in a modelling a complicated spectrum which also features: (i) a complex continuum heavily modified by signatures of substantial absorption/emission from ionized material $<2$\,keV, (ii) significant absorption features from highly-ionized Fe at $\sim$7--8\,keV, and (iii) an intricate Fe\,K emission complex at $\sim$6--7\,keV.  Furthermore, the effect of significant unmodelled residuals is magnified due the high statistical weight associated with an X-ray spectrum containing $\sim$2.6 $\times 10^{6}$\,counts.  Nevertheless, this is a phenomenological description and the fit with its associated residuals is shown in Fig.~\ref{fig:sed}.  We note that this parametrization is used as the ionizing SED with which we created the custom ionized emission / absorption grids used in \citep{Pounds15a, Pounds15b}, with an ionizing luminosity of $\sim$3.8 $\times 10^{45}$\,erg\,s$^{-1}$ from 1--1\,000\,Rydberg.  Such an approach is an improvement over the more standard method which typically consists of using grids generated with generic SEDs, which may not be appropriate for many sources.

\begin{figure}
\begin{center}
\rotatebox{0}{\includegraphics[width=8.4cm]{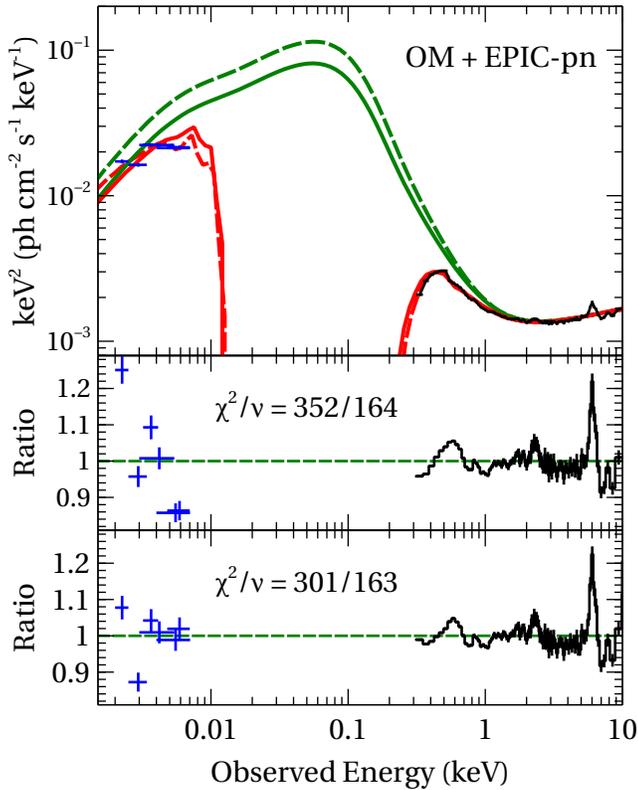}}
\end{center}
\vspace{-15pt}
\caption{Upper panel: the time-averaged OM (blue) + EPIC-pn (black) SED of PG\,1211+143 fitted with a model of the form: \textsc{redden} $\times$ \textsc{tbabs}$_{\rm Gal}$ $\times$ \textsc{tbabs} $\times$ \textsc{optxagnf} (red).  Middle panel: the data/model ratio of the residuals to this model.  Lower panel: the ratio of the residuals while allowing the reddening to be a free parameter.  The solid green line in the upper panel shows the intrinsic unabsorbed, de-reddened SED model.  The dashed lines in the upper panel show the model with the increased reddening.  See Section~\ref{sec:sed} for details.}
\label{fig:optxagnf}
\end{figure}

Finally, we also fit the optical--X-ray SED with the physical model, \textsc{optxagnf} \citep{Done12}, comprising: (i) a colour-temperature-corrected spectrum from an accretion disc; (ii) thermal Comptonization by optically-thick, low-temperature disc material, and (iii) thermal Comptonization by optically-thin, high-temperature coronal material.  The three components are fitted together self-consistently, assuming that they are all powered by the gravitational energy released due to accretion, with components (ii) and (iii) producing the soft X-ray excess $\lesssim 2$\,keV and harder power-law emission $\gtrsim 2$\,keV, respectively.

We fixed the comoving radial distance of the source at 331\,Mpc, the black hole spin to zero and the outer disc radius to $10^{5}$\,r$_{\rm g}$.  We constrained the black hole mass to lie in the range $M_{\rm BH} = 10^{7-8}$\,$M_{\odot}$ \citep{Kaspi00,Peterson04}.  The normalization was frozen at unity, as per the model requirement, since the flux is determined by the mass, Eddington ratio, spin and distance.  We also included two neutral absorbers (one fixed at the Galactic column, one free) and associated reddening [$E(B-V) = 0.035$], such that the model was of the form: \textsc{redden} $\times$ \textsc{tbabs}$_{\rm Gal}$ $\times$ \textsc{tbabs} $\times$ \textsc{optxagnf}.

Having fitted the model, we find a best-fitting power law slope of $\Gamma = 1.8$ with a resulting black hole accretion rate of $L / L_{\rm Edd} \sim 0.53$.  The black hole mass was $M_{\rm BH} \sim 8 \times 10^{7}$\,$M_{\odot}$.  The remaining interesting parameters consist of the coronal radius, $r_{\rm cor} \sim 15$\,$r_{\rm g}$, the electron temperature and optical depth of the soft Comptonizing material: $kT_{\rm e} \sim 0.8$\,keV; $\tau \sim 5$ and the fraction of the power law below $r_{\rm cor}$ which is emitted in the hard Comptonization component, $f_{\rm pl} \sim 0.2$.  Finally, in addition to the Galactic column of $3.06 \times 10^{20}$\,cm$^{-2}$, the additional neutral absorber requires a column density of $\sim$2 $\times 10^{20}$\,cm$^{-2}$.  From the fit we infer a bolometric luminosity of $L_{\rm bol} \sim 6 \times 10^{45}$\,erg\,s$^{-1}$.  The fit is shown in Fig.~\ref{fig:optxagnf} and, while statistically poor ($\chi^{2}/\nu = 352/164$), this is due to it being an oversimplification of a highly complex spectrum.  In particular, there is a significant contribution to the residuals from the Fe\,K band and the OM points, where it may be the case that the disc emission component rises more steeply than the OM points predict.  An improvement to the fit can be made by allowing the reddening to be a free parameter, which rises to $E(B-V) \sim 0.09$.  To compensate, the additional neutral absorber increases in column density from $\sim$2 $\times 10^{20}$ to $\sim$4 $\times 10^{20}$\,cm$^{-2}$ while the black hole mass increases slightly to $M_{\rm BH} \sim 9 \times 10^{7}$\,$M_{\odot}$.  The fit improves by $\Delta \chi^{2} = 51$ to $\chi^{2}/\nu = 301/163$ and predicts a higher bolometric luminosity of $L_{\rm bol} \sim 9 \times 10^{45}$\,erg\,s$^{-1}$.  The residuals are shown in the lower panel of Fig.~\ref{fig:optxagnf}.  We note that the fit is consistent with the {\it Swift} UVOT data in that the best-fitting parameters do not significantly change when the UVOT data are included.  

We also tried fixing the black hole mass at $M_{\rm BH} = 3 \times 10^{7}$\,$M_{\odot}$, more consistent with \citet{Kaspi00}.  In this instance, while there was no significant change to the majority of the free parameters, the accretion rate rose to $L / L_{\rm Edd} \gtrsim 2$ and the inferred bolometric luminosity fell to $L_{\rm bol} \sim 2 \times 10^{45}$\,erg\,s$^{-1}$.  The fit is still statistically poor with $\chi^{2}/\nu = 400/165$ and, interestingly, worse than the simple broken-power-law-SED fit (see above).  This may be because the soft X-rays and optical/UV points (which drive the fit but with uncertain systematics) are less well modelled.  Nevertheless, the model provides a useful illustrative account of the optical--X-ray SED.  A more detailed broad-band analysis will be presented in a forthcoming paper.

\subsubsection{{\it Swift} monitoring} \label{sec:swift_monitoring}

\begin{figure}
\begin{center}
\rotatebox{0}{\includegraphics[width=8.4cm]{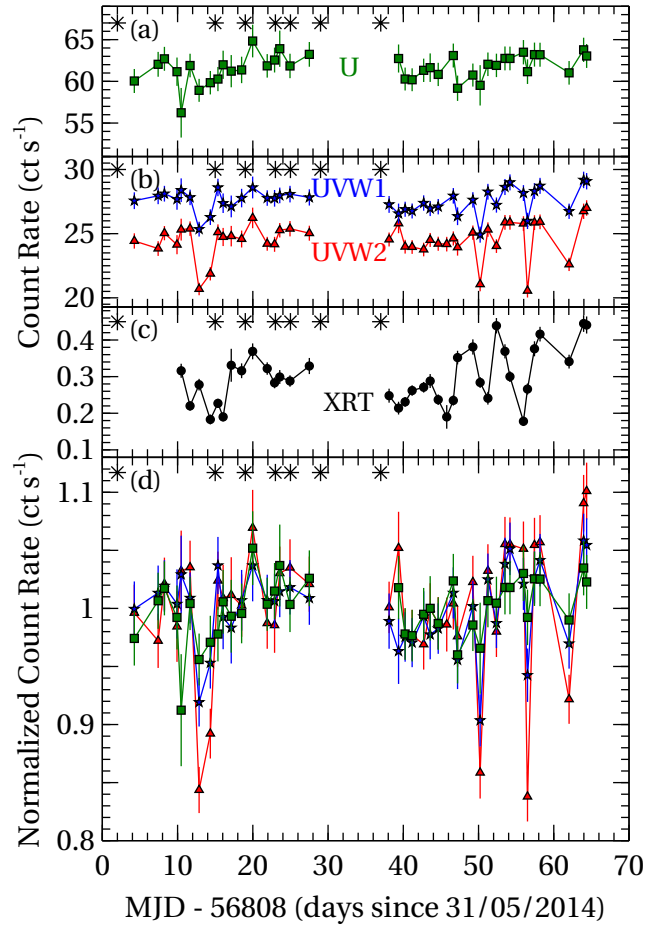}}
\end{center}
\vspace{-15pt}
\caption{The {\it Swift} UVOT/XRT light curve.  Panel a: the corrected count rates for the U (green crosses) filter; panel b: UVW1 (blue stars) and UVW2 (red triangles) filters; panel c: XRT (black circles); panel d: the overlaid U, UVW1 and UVW2 light curves, having each been normalized to their own mean.  The asterisks mark the times of the {\it XMM-Newton} observations.}
\label{fig:uvot_lc}
\end{figure}

The long-term {\it Swift} UVOT light curve is shown in Fig.~\ref{fig:uvot_lc} (panels a and b).  The mean count rates for the U, UVW1 and UVW2 filters are $61.64 \pm 0.26$, $27.58 \pm 0.11$ and $24.52 \pm 0.10$\,ct\,s$^{-1}$, respectively.  The amplitude of variability is similar to that found with the OM (Section~\ref{sec:optical_and_uv_monitoring}) with fractional variabilities, $F_{\rm var}$, of $\sim$0.5, $\sim$1.2 and $\sim$2.2\,per cent, respectively.  Panel d of Fig.~\ref{fig:uvot_lc} shows the three light curves overlaid, having each been normalized to their own mean.  The optical/UV variability appears to be weak although somewhat correlated with coincident peaks and troughs in the light curve while the data from the UVW2 filter appear to display the strongest fractional variability.

XRT counts were extracted using the online XRT products builder\footnote{\url{http://swift.ac.uk/user_objects/}} \citep{EvansBeardmorePage09}, which performs all necessary processing and provides background-subtracted light curves and spectra.  The 37 XRT observations vary in count rate by up to a factor of $\sim$3 over the $\sim$2-month period ($F_{\rm var} \sim 22.3$\,per cent) with a mean count rate of 0.296\,ct\,s$^{-1}$, ranging in observed flux from $F_{\rm 0.2-10} \sim 6 \times 10^{-12}$ to $F_{\rm 0.2-10} \sim 1.5 \times 10^{-11}$\,erg\,cm$^{-2}$\,s$^{-1}$ (mean flux, $F_{\rm 0.2-10} \sim 9.6 \times 10^{-12}$\,erg\,cm$^{-2}$\,s$^{-1}$).  We note that spectral shape of the time-averaged $\sim$47.1\,ks {\it Swift} XRT spectrum is consistent with the {\it XMM-Newton} EPIC-pn spectrum (Fig.~\ref{fig:2014_spectra}).  The XRT light curve is shown in Fig.~\ref{fig:uvot_lc} (panel c).

We now test for any correlation between wavelength bands on long timescales.  The cross-correlation function (CCF) is perhaps the most standard tool for measuring correlations between two time series (e.g. \citealt{BoxJenkins76}).  However, this requires that the time series be evenly sampled, while the sampling time between our {\it Swift} observations is highly uneven.  As such, we alleviate this by using the discrete correlation function (DCF; \citealt{EdelsonKrolik88}) to estimate the CCF.

We used the mean count rates for each UVOT exposure (typically a few hundred seconds in length), setting it at the middle of each exposure bin.  Likewise, we used the mean XRT count rates from all available snapshots and set them at the middle of each exposure bin (typically $\sim$0.5--1.5\,ks in length).  While the total duration of the {\it Swift} campaign is $\sim$60\,days, fewer pairs contribute to the DCF at longer lags, greatly reducing the certainty on the estimates.  As such, we computed the DCFs over the range -20 $<$ lag $<$ +20 days with a time bin size of 2 days, using the UVW2 filter as a reference band, and ensuring that there were at least 10 individual measurements per bin (in practice, there were typically $\sim$20--40 measurements per bin).

\begin{figure}
\begin{center}
\rotatebox{0}{\includegraphics[width=8.4cm]{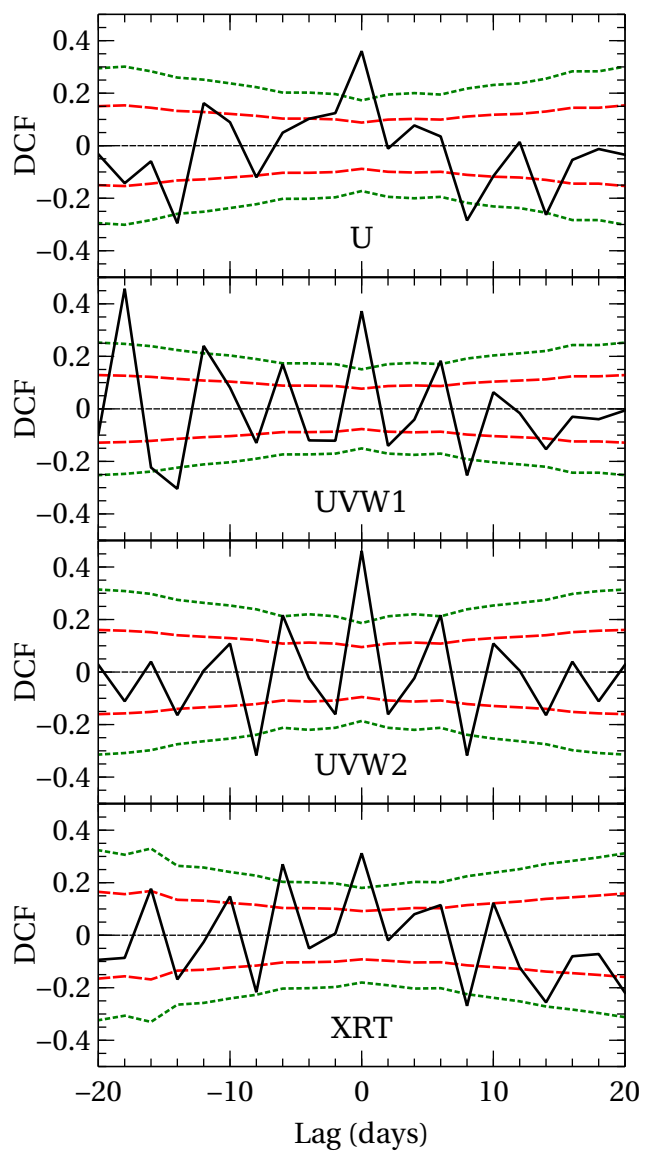}}
\end{center}
\vspace{-15pt}
\caption{The {\it Swift} DCFs for PG\,1211+143, comparing the U, UVW1, UVW2 and XRT (0.2--10\,keV) bands with the UVW2 band.  A positive lag means that the band in question lags behind the UVW2 band.  The red dashed and green dotted lines correspond to 1$\sigma$ (68.3\,per cent) and 95\,per cent confidence intervals, respectively.}
\label{fig:swift_dcf}
\end{figure}

We estimated rough confidence intervals using {\it Bartlett's formula} \citep{Bartlett55}; a method which is described in \citet{SmithVaughan07} and operates under the assumption that the data are stationary.  This involves computing the product of the two empirical autocorrelation functions (ACFs), using the same time delay bins as the DCF calculation, summed over all lags.  The 1$\sigma$ confidence band on the DCF is then: $\pm \sqrt{S/N}$, where $S$ is the total sum of the ACFs and $N$ is the number of pairs in each lag bin.  We also computed the 95\,per cent confidence region ($\pm 1.96\sigma$).  The DCFs are shown in Fig.~\ref{fig:swift_dcf}\footnote{We note that we do not weight the data by their uncertainties due to arising complications in interpeting the DCF amplitudes (see \citealt{WhitePeterson94}).}, where a positive lag denotes that the data are lagging behind the UVW2 data.  It can be seen that there is a significant correlation at zero lag between the U, UVW1 and XRT data with the UVW2 band, which rapidly falls off at longer lags.  There are additional peaks in the DCFs but they are largely within the confidence intervals and may exist as an artefact of the DCF binning.  We also computed the DCFs with the smallest possible time bin size (i.e. $< 1$\,day) and find a general improvement to the significance of the correlation at zero lag with the U, UVW1 and XRT bands having DCF values (compared to the UVW2 band) of 0.47, 0.59 and 0.34 with corresponding 95\,per cent confidence level estimates of 0.31, 0.30 and 0.23, respectively.

Finally, we note the presence of a couple of sharp dips in the UVOT light curve, particularly in the UVW1 and UVW2 bands.  These manifest themselves at MJD $\sim 56858$ and MJD $\sim 56864$ (see Fig.~\ref{fig:uvot_lc}).  Additionally, the UVOT count rate drops significantly over two consecutive pointings at MJD $\sim 56820$ (although this coincides with a known drop in X-ray flux in rev2659).  Such sharp dips may arise from regions of reduced UVOT detector sensitivity, as discussed by \citet{Gelbord15} and \citet{Edelson15}.  Having checked the raw UVOT images, we find that the source position during the observation at MJD $\sim 56864$ falls into one of the possible `bad' regions on the UVOT detector identified by \citet{Edelson15}.  While the remaining dips do not fall into such regions, it may be conceivable that other such previously unidentified patches of detector degradation do exist.  As such, we tested the robustness of our DCF measurements by recomputing them with the sharp dips removed from our analysis.  We find that the zero lag ($< 1$\,day) correlation is still significant with the U, UVW1 and XRT bands having DCF values of 0.50, 0.50 and 0.45, respectively.

\section{Discussion and conclusions} \label{sec:discussion_and_conclusions}

In this paper we have presented the average timing properties of a long $\sim$630\,ks {\it XMM-Newton} observation plus $\sim$5-week {\it Swift} monitoring campaign of PG\,1211+143.  Here, we discuss these results.

\subsection{The power spectrum} \label{sec:discussion_psd}

In Section~\ref{sec:psd} we modelled the broad-band 0.2--10\,keV PSD of PG\,1211+143 and showed that there is a well-detected bend, occurring at $\nu_{\rm b} = 6.9^{+2.4}_{-1.9} \times 10^{-5}$\,Hz.  This is the first detection of the high-frequency break in the PSD of this source.  While previous studies (e.g. \citealt{DeMarco11}; \citealt{Gonzalez-MartinVaughan12}) have analysed the PSD of PG\,1211+143, none had sufficient data on a long enough timescale to detect the bend and estimate its characteristic frequency.  

Based on simple scaling arguments, a somewhat linear relationship between the PSD bend time-scale and black hole mass might be expected (e.g. \citealt{Fender07}).  Indeed, an approximately linear correlation has been reported for a number of AGN (e.g. \citealt{UttleyMcHardyPapadakis02}; \citealt{Markowitz03}). The observed bend frequency for PG\,1211+143 corresponds to a time-scale of $T_{\rm b} = 1 / \nu_{\rm b} = 14^{+6}_{-3}$\,ks (or 0.16$^{+0.07}_{-0.03}$\,days).  \citet{Gonzalez-MartinVaughan12} provide a simple scaling relation between the observed bend time-scale and black hole mass for AGN: 

\begin{equation} \label{eq:bend_time-scale_vs_mass} {\rm log}(T_{\rm b}) = A{\rm log}(M_{\rm BH}) + C,  \end{equation}

where $T_{\rm b}$ is the break time-scale in days, $M_{\rm BH}$ is the black hole mass in units of $M_{\odot}$ and $A$ and $C$ are coefficients.  Assuming a black hole mass for PG\,1211+143 of $M_{\rm BH} \sim 10^{7-8}$\,$M_{\odot}$ \citep{Kaspi00,Peterson04} and values of $A = 1.09$ and $C = -1.70$, as derived by \citet{Gonzalez-MartinVaughan12} from fitting equation~\ref{eq:bend_time-scale_vs_mass} to a sample of known Seyferts, this predicts a bend time-scale in the range $\nu_{\rm b} \sim 0.2$--3\,days.  Given our observed value of $\nu_{\rm b} \sim 0.16$\,days, this is more consistent with a lower black hole mass estimate (e.g. $\sim$10$^{7}$\,$M_{\odot}$).

Additionally, the mass-timescale relation was extended by \citet{McHardy06} to include a dependence on bolometric luminosity, $L_{\rm bol}$ (also see \citealt{Koerding07}).  In this instance, equation~\ref{eq:bend_time-scale_vs_mass} contains an additional term: $B{\rm log}(L_{\rm bol})$, where $L_{\rm bol}$ is in units of $10^{44}$\,erg\,s$^{-1}$.  For this relation, \citet{Gonzalez-MartinVaughan12} derive best-fitting values of $A = 1.34$, $B = -0.24$ and $C = -1.88$.  For the case of PG\,1211+143, this predicts $\nu_{\rm b}$ to lie in the range $\sim$0.1--2.3\,days (assuming a value of $L_{\rm bol} \sim 6 \times 10^{45}$\,erg\,s$^{-1}$; \citealt{Gonzalez-MartinVaughan12}), also consistent with what we observe.

\subsection{The rms variability} \label{sec:rms-variability}

PG\,1211+143 displays enhanced variability when brighter (see Fig.~\ref{fig:rms-flux}).  This rms-flux relation is linear in the soft band, consistent with the rapid variability observed in accreting objects over a wide range in mass and luminosities such as XRBs, ULXs, AGN and CVs (e.g. \citealt{UttleyMcHardy01,Gleissner04,HeilVaughan10,Scaringi12}).  A natural explanation for a linear rms-flux relation is in terms of the propagating fluctuations model, whereby the X-ray source responds to variations in the mass accretion rate, which multiplicatively couple to one another as they propagate inwards to the inner disc (e.g. \citealt{UttleyMcHardyVaughan05}).  The X-ray variability of PG\,1211+143 may be complicated somewhat by the presence of a variable warm absorber / wind, which imprints significant features in the spectrum.  However, given that the rms-flux relation is consistent with that found in XRBs, we suggest that similar intrinsic accretion processes are at work here, albeit scaled according to the higher black hole mass, with the dominant short timescale variations arising from intrinsic process linked to the hard X-ray emitting source.

\subsection{The rms spectrum} \label{sec:discussion_rms-spectrum}

In Fig.~\ref{sec:rms-spectrum}, we explored the energy dependence of the rms variability, averaged across all orbits.  Over the `within-observation' frequency ranges, the fractional rms is roughly constant with energy; i.e. with, say, the flux at both 0.5 and 8\,keV varying by $\sim$10\,per cent rms on timescales of tens of ks.  As such, on these timescales, the hard and soft power laws are generally varying with nearly the same fractional amplitude\footnote{Conversely, if just, for example, the hard / soft power law was varying, we would see a significant drop in the rms at low / high energies where it becomes more diluted by the constant flux.}.  On longer timescales, however, higher-amplitude variations come into play as the soft band is observed to display enhanced variability with respect to the hard band.  Therefore, between observations, the relative strength of the soft and hard power laws is varying with large changes in the normalization of the soft component (i.e. the ``soft excess'') dominating the variability.  This is illustrated in Fig.~\ref{fig:difference_ratio_spectra}, where the difference spectrum is observed to be steep and can be fitted with a power law with a photon index of $\Gamma \sim 2.9$.  Additionally, this can also be observed in Fig.~\ref{fig:2014_spectra} where the 7 {\it XMM-Newton} time-averaged spectra are clearly more variable at low energies, where the soft component dominates.  If the steep, variable component of soft excess is due to Comptonization, one possible scenario may involve intrinsic coronal changes in terms of the electron temperature and/or optical depth of the corona, which may be cooler and/or thinner in the low-flux sequences where the soft component is weaker.  An alternative suggestion may be that the observed changes in the soft component are linked to variations in the warm absorber / outflow - in particular, we note that the soft component appears to be intrinsically weaker when the soft X-ray warm absorber is strongest (e.g. rev2659; see Fig.~\ref{fig:variable_absorption}).

Superimposed on these long timescale changes, we also observe additional structure in the fractional rms spectrum at $\sim$0.7--0.8\,keV, which can also be observed in the ratio spectrum (see Fig.~\ref{fig:difference_ratio_spectra}).  This additional structure is likely due to complex variations in the ionized absorption imprinted in the observed X-ray spectrum.  In particular, significant variations in the observed strength of the Fe UTA at $\sim$0.8\,keV can be seen between orbits (see Fig.~\ref{fig:variable_absorption}), which may perhaps arise from changes in the column density of a line-of-sight absorbing medium.  The details of the ionized absorbing material and its variability are discussed in a number of companion papers (i.e. \citealt{Pounds15a, Pounds15b, Lobban16}).  Finally, at $\sim$6--7\,keV, we also observe a characteristic drop in the rms which most likely arises from a quasi-constant component of Fe\,K emission, such as a narrow core of Fe\,K$\alpha$ emission originating in distant material (this component can be observed in the residuals in Figs~\ref{fig:sed} and~\ref{fig:optxagnf}).

\subsection{Optical/UV monitoring} \label{sec:discussion_optical_uv}

In Section~\ref{sec:optical_and_uv_monitoring}, we investigated the optical/UV properties of PG\,1211+143 using data acquired with the OM on-board {\it XMM-Newton} and $\sim$2\,months of contemporaneous ToO monitoring with {\it Swift}.  The within-observation UV variability is found to be small (e.g. Fig.~\ref{fig:uvw1_pn_lc}), while the between-observation optical/UV variability is found to be a few per cent ($F_{\rm var}$) on time-scales of $\sim$days.  The {\it XMM-Newton} OM versus EPIC-pn data suggest a peak in the optical/UV/X-ray flux around rev2663-2664 - however, we are unable to confirm any significant correlation, given the small number of observations (i.e. 7) and low UV variability amplitude.  Nevertheless, by computing the DCFs from the {\it Swift} monitoring campaign (Fig.~\ref{fig:swift_dcf}), we find that the optical/UV data are well correlated ($> 95$\,per cent level) with a lag $\lesssim \pm1$\,day (i.e. consistent with zero lag).  Meanwhile, there is a suggestion of a marginally significant correlation between the UV and X-ray light curves, again consistent with zero lag ($\lesssim \pm1$\,day).

Optical/UV variations correlated with X-rays on short time-scales have been observed in a number of type-1 Seyferts (e.g. MR\,2251-178: \citealt{Arevalo08}; Mrk 79: \citealt{Breedt09}; NGC 3783: \citealt{Arevalo09}; NGC 4051: \citealt{Breedt10}, \citealt{AlstonVaughanUttley13}; NGC 5548: \citealt{McHardy14}, \citealt{Edelson15}) where such variations are typically discussed in terms of a disc reprocessing model.  Here, the primary X-rays illuminate the accretion disc and are reprocessed, producing UV and optical emission with a time lag, $\tau$, determined by the light travel time to the corresponding emission site.  As such, large X-ray fluctuations may result in smaller-amplitude wavelength-dependent time delays where, in the case of an optically thick disc, the longer-wavelength counterparts are delayed with the expected relation: $\tau \propto \lambda^{4/3}$ (e.g. \citealt{CackettHorneWinkler07}).  While the bulk of the observed optical/UV luminosity may be ``intrinsic'' (i.e. due to internal viscous heating), this should remain constant on long time-scales while an additional fraction of observed light may vary due to X-ray reprocessing.

In the case of PG\,1211+143, the X-ray luminosity is observed to vary considerably.  If these X-rays then illuminate a standard optically thick accretion disc (assuming $\alpha \sim 0.1$, $H/R \sim 0.01$ and a central compact X-ray source at $\sim$6\,$r_{\rm g}$ above the mid-plane) and assuming a black hole mass of $10^{7}$\,$M_{\odot}$ and $L/L_{\rm Edd} \sim 0.5$, the U-band (effective wavelength: 3\,440\,\AA) would have an emission-weighted radius of $\sim$500\,$r_{\rm g}$ (see equations 3.20 and 3.21 of \citealt{Peterson97}), corresponding to a light-crossing time of $\sim$0.3\,days\footnote{Taking the effective wavelengths of each OM filter (see Table~\ref{tab:om_count_rates}) gives expected time delays in the range $\tau \sim 0.15$--$0.5$\,days, all of which are smaller than our sampling rate.}.  As such, X-ray reprocessing would predict an optical/UV/X-ray correlation with no obvious delays, given the resolution of our data.  Contrarily, the viscous timescale is on the order of $\sim$2 $\times 10^{3}$\,years.  A reprocessing scenario is consistent with the absolute change in optical / X-ray luminosities where the X-ray variations are a factor of $\sim$20 larger but are also $\sim$10--15 times weaker in $L$.  As such, the large variations in the weak X-rays are enough to drive the small variations in the large optical luminosity.

This may be consistent with the analysis of \citet{Bachev09} who undertook a multi-wavelength monitoring campaign of PG\,1211+143 using {\it Swift} and ground-based facilities.  Similar to our results here, they found only minor optical/UV flux changes relative to the large observed X-ray variations.  However, their inter-band interpolated cross-correlation analysis did suggest the presence of wavelength-dependent time delays, which may indicate that at least some fraction of the X-rays is reprocessed into longer-wavelength emission.  Given the hint of possible optical/UV/X-ray correlations, we note that PG\,1211+143 may be a suitable target for a more comprehensive monitoring campaign with {\it Swift}, sampling the source more extensively in time and flux.

\section*{Acknowledgements}

This research has made use of the NASA Astronomical Data System (ADS), the NASA Extragalactic Database (NED) and is based on observations obtained with {\it XMM-Newton}, an ESA science mission with instruments and contributions directly funded by ESA Member States and NASA.  This research is also based on observations with the NASA/UKSA/ASI mission {\it Swift}. AL, SV and JR acknowledge support from the UK STFC.

\label{lastpage}


\begin{thebibliography}{}

\bibitem[\protect\citeauthoryear{Alston, Vaughan \& Uttley}{Alston et al.}{2013}]{AlstonVaughanUttley13}Alston W. N., Vaughan S., Uttley P., MNRAS, 435, 1511

\bibitem[\protect\citeauthoryear{Alston, Done \& Vaughan}{Alston et al.}{2014}]{AlstonDoneVaughan14}Alston W. N., Done C., Vaughan S., 2014, MNRAS, 439, 1548

\bibitem[\protect\citeauthoryear{Ar\'{e}valo et al.}{2008}]{Arevalo08}Ar\'{e}valo P., Uttley P., Kaspi S., Breedt E., Lira P., McHardy I. M., 2008, MNRAS, 389, 1479

\bibitem[\protect\citeauthoryear{Ar\'{e}valo et al.}{2009}]{Arevalo09}Ar\'{e}valo P., Uttley P., Lira P., Breedt E., McHardy I. M., Churazov E., 2009, MNRAS, 397, 2004

\bibitem[\protect\citeauthoryear{Arnaud}{1996}]{Arnaud96}Arnaud K. A., 1996, in Jacoby G. H., Barnes J., eds, ASP Conf. Ser. Vol. 101, Astronomical Data Analysis Software and Systems V. Astron. Soc. Pac., San Francisco, p. 17

\bibitem[\protect\citeauthoryear{Bachev et al.}{2009}]{Bachev09}Bachev R., Grupe D., Boeva S., Ovcharov E., Valcheva A., Semkov E., Georgiev Ts., Gallo L. C., 2009, MNRAS, 399, 750

\bibitem[\protect\citeauthoryear{Barret \& Vaughan}{2012}]{BarretVaughan12}Barret D., Vaughan S., 2012, ApJ, 746, 131

\bibitem[\protect\citeauthoryear{Bartlett}{1955}]{Bartlett55}Bartlett M. S., 1955, An Introduction to Stochastic Processes (CUP: Cambridge)

\bibitem[\protect\citeauthoryear{Box \& Jenkins}{1976}]{BoxJenkins76}Box G. E. P., \& Jenkins G. M., eds, 1976, Time Series Analysis: Forecasting and Control

\bibitem[\protect\citeauthoryear{Breedt et al.}{2009}]{Breedt09}Breedt E., Ar\'{e}valo P., McHardy I. M., et al., MNRAS, 394, 427

\bibitem[\protect\citeauthoryear{Breedt et al.}{2010}]{Breedt10}Breedt E., McHardy I. M., Ar\'{e}valo P., et al., MNRAS, 403, 605

\bibitem[\protect\citeauthoryear{Burrows et al.}{2005}]{Burrows05}Burrows D. N., et al., 2005, SSRv, 120, 165

\bibitem[\protect\citeauthoryear{Cackett, Horne \& Winker}{Cackett et al.}{2007}]{CackettHorneWinkler07}Cackett E. M., Horne K., Winkler H., 2007, MNRAS, 380, 669

\bibitem[\protect\citeauthoryear{Cameron et al.}{2012}]{Cameron12}Cameron D. T., McHardy I. M., Dwelly T., Breedt E., Uttley P., Lira P., Ar\'{e}valo P., 2012, MNRAS, 422, 902

\bibitem[\protect\citeauthoryear{Cardelli, Clayton \& Mathis}{1989}]{CardelliClaytonMathis89}Cardelli J. A., Clayton G. C., Mathis J. S., 1989, ApJ, 345, 245

\bibitem[\protect\citeauthoryear{Collier et al.}{1998}]{Collier98}Collier S. J., Horne K., Kaspi S., et al., 1998, ApJ, 500, 162

\bibitem[\protect\citeauthoryear{Collin}{2001}]{Collin01}Collin S., 2001, in Aretxaga I., Kunth D., M\'{u}jica R., eds, Advanced Lectures on the Starbust-AGN Accretion and Emission Processes in AGN. p. 167

\bibitem[\protect\citeauthoryear{Cui et al.}{1997}]{Cui97}Cui W., Zhang, S. N., Focke, W., Swank, J. H., 1997, ApJ, 484, 383

\bibitem[\protect\citeauthoryear{De Marco et al.}{2011}]{DeMarco11}De Marco B., Ponti G., Uttley P., Cappi M., Dadina M., Fabian A. C., Miniutti G., 2011, MNRAS, 417, 98

\bibitem[\protect\citeauthoryear{den Herder et al.}{2001}]{denHerder01}den Herder J. W. et al., 2001, A\&A, 365, 7

\bibitem[\protect\citeauthoryear{Done et al.}{2012}]{Done12}Done C., Davis S. W., Jin C., Blaes O., Ward M., 2012, MNRAS, 420, 1848

\bibitem[\protect\citeauthoryear{Edelson \& Krolik}{1988}]{EdelsonKrolik88}Edelson R. A., Krolik J. H., 1988, ApJ, 333, 646

\bibitem[\protect\citeauthoryear{Edelson et al.}{2015}]{Edelson15}Edelson R., Gelbord J. M., Horne K., et al., 2015, preprint (arXiv:1501.05951)

\bibitem[\protect\citeauthoryear{Evans, Beardmore \& Page}{2009}]{EvansBeardmorePage09}Evans P. A., Beardmore A. P., Page K. L., 2009, MNRAS, 397, 1177

\bibitem[\protect\citeauthoryear{Fender et al.}{2007}]{Fender07}Fender R., Koerding E., Belloni T., Uttley P., McHardy I., Tzioumis T., 2007, preprint (arXiv:0706.3838)

\bibitem[\protect\citeauthoryear{Gaskell}{2004}]{Gaskell04}Gaskell M., 2004, ApJL, 612, L21

\bibitem[\protect\citeauthoryear{Gehrels et al.}{2004}]{Gehrels04}Gehrels N., et al., 2004, ApJ, 611, 1005

\bibitem[\protect\citeauthoryear{Gelbord et al.}{2015}]{Gelbord15}Gelbord J., Gronwall C., Grupe D., Vanden Berk D., Wu J., 2015, preprint (arXiv:1505.05248)

\bibitem[\protect\citeauthoryear{George \& Fabian}{1991}]{GeorgeFabian91}George I., Fabian A. C., 1991, MNRAS, 249, 352

\bibitem[\protect\citeauthoryear{Gleissner et al.}{2004}]{Gleissner04}Gleissner T., Wilms J., Pooley G. G., et al., 2004, A\&A, 425, 1061

\bibitem[\protect\citeauthoryear{Gonz{\'a}lez-Mart{\'i}n \& Vaughan}{2012}]{Gonzalez-MartinVaughan12}Gonz{\'a}lez-Mart{\'i}n O., Vaughan S., 2012, A\&A, 544, 80

\bibitem[\protect\citeauthoryear{Guilbert \& Rees}{1988}]{GuilbertRees88}Guilbert P. W., Rees M. J., 1988, MNRAS, 233, 475

\bibitem[\protect\citeauthoryear{Haardt \& Maraschi}{1991}]{HaardtMaraschi91}Haardt F., Maraschi I., 1991, ApJL, 380, L51

\bibitem[\protect\citeauthoryear{Haardt \& Maraschi}{1993}]{HaardtMaraschi93}Haardt F., Maraschi I., 1993, ApJ, 413, 507

\bibitem[\protect\citeauthoryear{Heil \& Vaughan}{2010}]{HeilVaughan10}Heil L. M., Vaughan S., 2010, MNRAS, 405, 86

\bibitem[\protect\citeauthoryear{Jansen et al.}{2001}]{Jansen01}Jansen F. et al., 2001, A\&A, 365, 1

\bibitem[\protect\citeauthoryear{Kalberla et al.}{2005}]{Kalberla05}Kalberla P. M. W., Burton W. B., Hartmann D., Arnal E. M., Bajaja E., Morras R., P\"{o}ppel W. G. L., 2005, A\&A, 440, 775

\bibitem[\protect\citeauthoryear{Kara et al.}{2014}]{Kara14}Kara E., Cackett E. M., Fabian A. C., Reynolds C., Uttley P., 2014, MNRAS, 439, L26

\bibitem[\protect\citeauthoryear{Kaspi et al.}{2000}]{Kaspi00}Kaspi S., Smith P. S., Netzer H., Maoz D., Jannuzi B. T., Giveon U., 2000, ApJ, 533, 631

\bibitem[\protect\citeauthoryear{K\"{o}rding et al.}{2007}]{Koerding07}K\"{o}rding E. G., Migliari S., Fender R., Belloni T., Knigge C., McHardy I., 2007, MNRAS, 380, 301 

\bibitem[\protect\citeauthoryear{Lawrence et al.}{1987}]{Lawrence87}Lawrence A., Watson M. G., Pounds K. A., Elvis M., 1987, Nature, 325,694

\bibitem[\protect\citeauthoryear{Lobban et al.}{2015}]{Lobban15}Lobban A. P., Vaughan S., Pounds K. A., Reeves J. N., 2015, in prep.

\bibitem[\protect\citeauthoryear{Lobban et al.}{2016}]{Lobban16}Lobban A. P., Vaughan S., Pounds K. A., Reeves J. N., 2016, in prep.

\bibitem[\protect\citeauthoryear{Markowitz et al.}{2003}]{Markowitz03}Markowitz A., Edelson R., Vaughan S., et al., 2003, ApJ, 593, 96

\bibitem[\protect\citeauthoryear{Marziani et al.}{1996}]{Marziani96}Marziani P., Sulentic J. W., Dultzin-Hacyan D., Calvani M., Moles M., 1996, ApJS, 104, 37

\bibitem[\protect\citeauthoryear{Mason et al.}{2001}]{Mason01}Mason K. O. et al., 2001, A\&A, 365, 36

\bibitem[\protect\citeauthoryear{McHardy et al.}{2004}]{McHardy04}McHardy I. M., Papadakis I. E., Uttley P., Page M. J., Mason K. O., 2004, MNRAS, 348, 783

\bibitem[\protect\citeauthoryear{McHardy et al.}{2006}]{McHardy06}McHardy I. M., Koerding E., Knigge C., Uttley P., Fender R. P., 2006, Nature, 444, 730

\bibitem[\protect\citeauthoryear{McHardy et al.}{2007}]{McHardy07}McHardy I. M., Ar\'{e}valo P., Uttley P., Papadakis I. E., Summons D. P., Brinkmann W., Page M. J., 2007, MNRAS, 382, 985

\bibitem[\protect\citeauthoryear{McHardy et al.}{2014}]{McHardy14}McHardy I. M., Cameron D. T., Dwelly T., et al., 2014, MNRAS, 444, 1469

\bibitem[\protect\citeauthoryear{Miller et al.}{2010}]{Miller10}Miller L., Turner T. J., Reeves J. N., Braito V., 2010, MNRAS, 408, 1928

\bibitem[\protect\citeauthoryear{Miyamoto \& Kitamoto}{1989}]{MiyamotoKitamoto89}Miyamoto S., Kitamoto S., 1989, Nature, 342, 773

\bibitem[\protect\citeauthoryear{Mushotzky, Done \& Pounds}{Mushotzky et al.}{1993}]{MushotzkyDonePounds93}Mushotzky R. F., Done C., Pounds K. A., 1993, ARA\@A, 31, 717

\bibitem[\protect\citeauthoryear{Nandra}{1998}]{Nandra98}Nandra K., Clavel J., Edelson R. A., Mushotzky R. F., Peterson B. M., Turner T. J., 2000, ApJ, 544, 734

\bibitem[\protect\citeauthoryear{Nowak}{2005}]{Nowak05}Nowak M., 2005, Ap\&SS, 300, 159

\bibitem[\protect\citeauthoryear{Papadakis, Nandra \& Kazanas}{Papadakis et al.}{2001}]{PapadakisNandraKazanas01}Papadakis I. E., Nandra K., Kazanas D., 2001, ApJ, 554, 133

\bibitem[\protect\citeauthoryear{Percival \& Walden}{1993}]{PercivalWalden93}Percival D. B., \& Walden A. T., 1993, Spectral analysis for physical applications: multi taper and conventional univariate techniques, Cambridge University Press, Cambridge

\bibitem[\protect\citeauthoryear{Peterson}{2007}]{Peterson97}Peterson B. M., 1997, An Introduction to Active Galactic Nuclei (Cambridge University Press: Cambridge)

\bibitem[\protect\citeauthoryear{Peterson et al.}{2004}]{Peterson04}Peterson B., Ferrarese L., Gilbert K. M., 2004, ApJ, 613, 682

\bibitem[\protect\citeauthoryear{Priestly}{1981}]{Priestly81}Priestly M. B., 1981, Spectral Analysis and Time Series, Academic Press, London

\bibitem[\protect\citeauthoryear{Pounds et al.}{2003}]{Pounds03}Pounds K. A., Reeves J. N., King A. R., Page K. L., O'Brien P. T., Turner M. J. L., 2003, MNRAS, 345, 705

\bibitem[\protect\citeauthoryear{Pounds \& Page}{2006}]{PoundsPage06}Pounds K. A., Page K. L., 2006, MNRAS, 360, 1123

\bibitem[\protect\citeauthoryear{Pounds \& Reeves}{2009}]{PoundsReeves09}Pounds K. A., Reeves J. N., 2009, MNRAS, 397, 249

\bibitem[\protect\citeauthoryear{Pounds et al.}{2015a}]{Pounds15a}Pounds K. A., Lobban A. P., Reeves J. N., Vaughan S., 2015a, submitted

\bibitem[\protect\citeauthoryear{Pounds et al.}{2015b}]{Pounds15b}Pounds K. A., Lobban A. P., Reeves J. N., Vaughan S., 2015b, submitted

\bibitem[\protect\citeauthoryear{Roming et al.}{2005}]{Roming05}Roming P. W. A., et al., 2005, SSRv, 120, 95

\bibitem[\protect\citeauthoryear{Scaringi et al.}{2012}]{Scaringi12}Scaringi S., Koerding E., Uttley P., Knigge C., Groot P. J., Still M., 2012, MNRAS, 421, 2854

\bibitem[\protect\citeauthoryear{Scott, Stewart \& Mateos}{2012}]{ScottStewartMateos12}Scott A., Stewart G., Mateos S., 2012, MNRAS, 412, 2633

\bibitem[\protect\citeauthoryear{Shakura \& Sunyaev}{1973}]{ShakuraSunyaev73}Shakura N. I., Sunyaev R. A., 1973, A\&A, 24, 337

\bibitem[\protect\citeauthoryear{Smith \& Vaughan}{2007}]{SmithVaughan07}Smith R., Vaughan S., 2007, MNRAS, 375, 1479

\bibitem[\protect\citeauthoryear{Uttley \& McHardy}{2001}]{UttleyMcHardy01}Uttley P., McHardy I. M., MNRAS, 2001, 323, 26

\bibitem[\protect\citeauthoryear{Uttley, McHardy \& Papadakis}{2002}]{UttleyMcHardyPapadakis02}Uttley P., McHardy I. M., Papadakis I. E., 2002, MNRAS, 332, 231

\bibitem[\protect\citeauthoryear{Uttley, McHardy \& Vaughan}{Uttley et al.}{2005}]{UttleyMcHardyVaughan05}Uttley P., McHardy I. M., Vaughan S., 2005, MNRAS, 359, 345

\bibitem[\protect\citeauthoryear{Vaughan}{2010}]{Vaughan10}Vaughan S., 2010, MNRAS, 402, 307

\bibitem[\protect\citeauthoryear{Vaughan et al.}{2003}]{Vaughan03}Vaughan S., Edelson R., Warwick R. S., Uttley P., 2003, MNRAS, 345, 1271

\bibitem[\protect\citeauthoryear{Vaughan, Fabian \& Nandra}{2003}]{VaughanFabianNandra03}Vaughan S., Fabian A. C., Nandra K., 2003, MNRAS, 339, 1237

\bibitem[\protect\citeauthoryear{Verner et al.}{1996}]{Verner96}Verner D. A., Ferland G. J., Korista K. T., Yakovlev D. G., 1996, ApJ, 465, 487

\bibitem[\protect\citeauthoryear{White \& Peterson}{1994}]{WhitePeterson94}White R. J., Peterson B. M., 1994, PASP, 106, 879

\bibitem[\protect\citeauthoryear{Willingale et al.}{2013}]{Willingale13}Willingale R., Starling R. L. C., Beardmore A. P., Tanvir N. R., O'Brien P. T., 2013, MNRAS, 431, 394

\bibitem[\protect\citeauthoryear{Wilms, Allen \& McCray}{Wilms et al.}{2000}]{WilmsAllenMcCray00}Wilms J., Allen A., McCray R., 2000, ApJ, 542, 914

\end{thebibliography}
\end{document}